\documentclass[preprint,12pt]{elsarticle}
\usepackage{slashed}
\usepackage{amssymb}
\usepackage{amsmath}
\usepackage{url}
\journal{Physics of Dark Universe}

\begin{document}

\begin{frontmatter}

%% Title, authors and addresses

%% use the tnoteref command within \title for footnotes;
%% use the tnotetext command for theassociated footnote;
%% use the fnref command within \author or \affiliation for footnotes;
%% use the fntext command for theassociated footnote;
%% use the corref command within \author for corresponding author footnotes;
%% use the cortext command for theassociated footnote;
%% use the ead command for the email address,
%% and the form \ead[url] for the home page:
%% \title{Title\tnoteref{label1}}
%% \tnotetext[label1]{}
%% \author{Name\corref{cor1}\fnref{label2}}
%% \ead{email address}
%% \ead[url]{home page}
%% \fntext[label2]{}
%% \cortext[cor1]{}
%% \affiliation{organization={},
%%             addressline={},
%%             city={},
%%             postcode={},
%%             state={},
%%             country={}}
%% \fntext[label3]{}

\title{F-mode Oscillations of Neutron Stars with Dark Matter from Neutron Decay: Implications for Gravitational-Wave Detectability}

%% use optional labels to link authors explicitly to addresses:
%% \author[label1,label2]{}
%% \affiliation[label1]{organization={},
%%             addressline={},
%%             city={},
%%             postcode={},
%%             state={},
%%             country={}}
%%
%% \affiliation[label2]{organization={},
%%             addressline={},
%%             city={},
%%             postcode={},
%%             state={},
%%             country={}}

\author{Wasif Husain} %% Author name
\ead{whusain@usc.edu.au}
%% Author affiliation
\affiliation{organization={School of Science, Technology and Engineering, University of Sunshine Coast, },%Department and Organization
            addressline={182, Victoria Square}, 
            city={Adelaide},
            postcode={5000}, 
            state={SA},
            country={Australia}}

%% Abstract
\begin{abstract}
%% Text of abstract
In this study, the impact of neutron decay into dark matter and various dark matter self-interaction strengths on neutron star properties have been explored. Using the quark-meson coupling (QMC) model for nucleon-only equations of state (EoSs), the effects of different matter compositions have been compared, including strange matter and self-interacting dark matter. The results demonstrate that increasing DM-DM self-repulsion stiffens the EoS, influencing the mass-radius relationship and stability of neutron stars. \par
Furthermore, fundamental mode (f-mode) oscillations have been analyzed, which serve as a diagnostic tool for probing neutron star interiors. The f-mode frequencies follow universal relations, reinforcing their applicability for constraining dense matter properties. It has been shown that neutron stars composed of nucleons-only and self-interacting dark matter exhibit a universal behavior in damping time and angular frequency, whereas strange matter and non-self-interacting dark matter deviate from this trend.\par
Importantly, it has been shown that for a GW energy release of $E \sim 10^{52}$ erg and a source distance of 25 Mpc, the characteristic strain and signal-to-noise ratio exceed the ET-D sensitivity threshold below $\sim$2.1 kHz for all models except the non-interacting DM case, demonstrating that neutron-to-dark matter decay scenarios, including the role of DM self-interactions, can be tested through next-generation gravitational-wave asteroseismology, offering a new probe of DM physics and the neutron lifetime anomaly.
\end{abstract}

%% Keywords
\begin{keyword}
%% keywords here, in the form: keyword \sep keyword
Neutron stars \sep Dark Matter \sep Neutron decay \sep f-mode oscillations.
%% PACS codes here, in the form: \PACS code \sep code

%% MSC codes here, in the form: \MSC code \sep code
%% or \MSC[2008] code \sep code (2000 is the default)

\end{keyword}

\end{frontmatter}

%% Add \usepackage{lineno} before \begin{document} and uncomment 
%% following line to enable line numbers
%% \linenumbers

%% main text
%%

%% Use \section commands to start a section
\flushbottom

\section{Introduction}
\label{sec:intro}
The hypothesis of self-interacting dark matter (SIDM) has gained prominence due to its potential to address discrepancies within the standard collisionless cold dark matter (CCDM) paradigm, alongside its theoretical plausibility within extensions of the Standard Model. 
Specifically, the CCDM model faces challenges such as the observed flat density profiles in dwarf galaxy cores, contrasting with the cuspy profiles predicted by simulations \cite{Moore_1999,Navarro_1997}, the overabundance of simulated satellite galaxies compared to observed counts \cite{Moore_1999,Klypin_1999,10.1093/mnras/264.1.201}, and  the "too big to fail" problem, wherein simulations predict massive, star-forming dwarf galaxies that remain undetected \cite{10.1111/j.1745-3933.2011.01074.x,Kanehisa_2024}.   

While baryonic interactions and statistical fluctuations have been proposed as potential resolutions to galactic anomalies, SIDM offers a compelling alternative capable of addressing these issues, and the neutron decay hypothesis provides a specific mechanism for dark matter production. Enhanced self-scattering rates in high-density regions, resulting from SIDM, could effectively flatten dwarf galaxy cores. The required self-interaction cross-sections, within the range of 0.1cm$^2/g < \sigma_{\chi\chi}/m_\chi <$10 cm$^2/g$, where $\sigma_{\chi\chi}$ is the self-interaction cross-section and $m_\chi$
the dark matter particle mass are, have been extensively studied \cite{10.1093/mnras/sts514,10.1093/mnrasl/sls053,10.1093/mnras/sts535,PhysRevLett.84.3760,ZHANG2023106967}.

Additionally, the neutron lifetime anomaly, where beam experiments yield longer lifetimes than bottle experiments \cite{Serebrov:2017bzo,Pattie:2017vsj,TAN2019134921,PhysRevC.85.065503}, suggests a potential decay channel into undetected dark sector particles \cite{Fornal:2018eol,Grinstein:2018ptl,Fornal:2020gto,RevModPhys.83.1173}. This has motivated the exploration of neutron decay into dark matter as a potential resolution. The neutron decay scenario, proposed by Fornal and Grinstein \cite{Fornal:2020gto}, posits that neutrons decay into a dark fermion ($\chi$), potentially explaining the observed neutron lifetime discrepancy. This hypothesis implies that neutron stars, composed primarily of neutrons, could serve as natural laboratories for studying dark matter interactions.  

The fundamental mode (f-mode) is a type of non-radial oscillation in compact stars, primarily governed by the star’s bulk properties such as mass, radius, and internal composition. These oscillations generate gravitational waves, and their frequency and damping time are sensitive to the equation of state (EoS) of the stellar interior. As such, the f-mode serves as a powerful diagnostic tool for probing the internal structure and composition of neutron stars and, more generally, for constraining the nature of dense matter under extreme conditions.

There are several promising avenues for measuring f-mode \cite{PhysRevD.65.063006,2010PhRvC..82b5803J,2014CQGra..31o5002V,2022PhRvD.106l3002Z} characteristics in compact stars. For instance, third-generation gravitational wave detectors like the Einstein Telescope \cite{Punturo_2010} and Cosmic Explorer \cite{Abbott_2017} are expected to achieve the sensitivity required for detecting f-mode signals from astrophysical sources \cite{PhysRevD.85.024030}.% Even with current instrumentation, estimates of f-mode frequencies can be obtained using the f-Love relations [64], which connect the f-mode frequency to the tidal deformability ($\Lambda$)—a parameter measurable during the late inspiral phase of binary neutron star mergers [65, 66].

Furthermore, the damping time and frequency of the f-mode are also linked to the star’s properties, such as mass, compactness, and moment of inertia, quantities that could be inferred from spin-orbit coupling effects in binary pulsar systems \cite{ASCENZI2024102935}. These interdependencies allow the f-mode to act as a bridge between observable gravitational wave signatures and the otherwise inaccessible interior physics of compact stars.
In this work, previous studies of neutron decay into dark matter have been extended by incorporating both nucleonic and strange matter EoS within the quark–meson coupling (QMC) model, and by explicitly evaluating the detectability of f-mode oscillations with next-generation gravitational-wave detectors. While earlier works have addressed stability and mass-radius constraints for such stars, a systematic investigation of characteristic strain and signal-to-noise ratios (SNR) in the context of ET-D sensitivity has not yet been carried out. These results therefore provide the first quantitative assessment of whether gravitational-wave observations of f-modes can constrain dark matter self-interactions and the presence of exotic matter in neutron stars.

This paper presents, for the first time, a systematic test of the neutron–to–dark matter decay model using f-mode oscillations and signal-to-noise ratio estimates. The study is organized as follows. Section 2 introduces the theoretical framework for neutron decay into dark matter and discusses astrophysical constraints. Section 3 provides an overview of neutron star formation, structure, and observed phenomena, while Section 4 addresses the equations of state governing ultradense matter. Section 5 outlines the stellar structure equations, and Section 6 develops advanced considerations of the EoS. Section 7 examines universal relations of f-mode oscillations, with subsections devoted to damping times (7.1) and mass-scaled angular frequencies (7.2). Section 8 compares f-mode frequencies predicted by nucleonic, strange matter, and dark matter models. Section 9 evaluates detectability by confronting the predicted f-mode signals with the Einstein Telescope sensitivity curve and presenting signal-to-noise ratios. Section 10 concludes with a summary of the main results and their implications for neutron star physics, gravitational-wave astronomy, and the search for dark matter signatures.

\section{Neutron decay into dark matter}
There remains an ongoing discrepancy between neutron lifetime measurements obtained using beam and bottle methods. Beam experiments, which detect proton decay, consistently yield lifetimes approximately 1\% longer than those measured in bottle experiments, where the decay products are not directly observed. Advances in ultracold neutron studies have refined bottle measurements, yet beam experiments continue to produce lifetimes that remain about eight seconds longer \cite{PAUL2009157,RevModPhys.83.1173,PhysRevLett.127.162501,10.1093/ptep/ptaa169,PhysRevLett.111.222501,Otono:2016fsv,Olive_2016,UCNt:2021pcg,doi:10.1126/science.aan8895,PhysRevC.97.055503,doi:10.1126/science.aan8895}. 
%Beam experiments, which detect proton decay, consistently yield lifetimes about 1\% longer than those measured in bottle experiments, where the decay products are not directly observed. The ultracold neutron studies have improved bottle measurements, reporting lifetimes around 877.7 seconds, whereas beam experiments continue to produce values close to 887.7 seconds \cite{PAUL2009157,RevModPhys.83.1173,PhysRevLett.127.162501,10.1093/ptep/ptaa169}. This difference amounts to a 4$\sigma$ deviation, approximately 8 seconds, highlighting a significant unresolved inconsistency.

%A persistent discrepancy exists between neutron lifetime measurements obtained via beam and bottle methods [1–8]. Specifically, beam experiments, which detect proton decay, yield a lifetime approximately 1\% longer than bottle experiments, where the decay products remain undetermined. Recent ultra-cold neutron (UCN) experiments have refined bottle measurements, reporting lifetimes around 877.7 seconds [9–11], while beam measurements consistently produce values near 887.7 seconds [10], resulting in a significant 4$\sigma$ (approximately 8 seconds) divergence.

To address this anomaly, Fornal and Grinstein \cite{Fornal:2018eol,Grinstein:2018ptl,Fornal:2020gto} proposed a dark decay channel wherein neutrons decay into dark sector particles, 
\begin{equation*}
    n \xrightarrow{} \chi +\phi
\end{equation*} 
including a dark fermion ($\chi$), alongside the standard 
mode. 
%\begin{equation*}
%    n \xrightarrow{} p + e^- + \bar{\nu}_e
%\end{equation*}

This hypothesis posits that weakly interacting dark matter particles remain undetected in beam experiments but contribute to the total decay rate in bottle experiments. This alternative to neutron-mirror neutron oscillation \cite{Fornal:2018eol,Grinstein:2018ptl} necessitates a dark fermion with a mass (m$_\chi$) nearly degenerate with the neutron within the range \( 937.9 \, \text{MeV} < m_\chi < 938.7 \, \text{MeV} \), more details of the model can be found in Ref.  \cite{Fornal:2018eol,Grinstein:2018ptl}. The accompanying boson cannot be a photon as shown in Ref. ~\cite{Tang:2018eln,Serebrov:2007gw}.\par
Furthermore, this dark decay hypothesis has been suggested as a potential resolution to the reactor anti-neutrino anomaly \cite{Serebrov:2007gw}. Notably, it imposes constraints on neutron star properties. If these dark fermions are non-interacting, the maximum neutron star mass is predicted to be below 0.7 M$_\odot$ \cite{Motta:2018rxp,Motta:2018bil,PhysRevLett.121.061801,2018_sa,Husain:2022bxl}, contradicting observed neutron star masses exceeding 2 M$_\odot$ \cite{2019ApJ...887L..24M,
2019ApJ...887L..21R,2018PhRvL.121p1101A,2021ApJ...918L..28M}. Therefore, a repulsive self-interaction among dark fermions is required to reconcile this discrepancy.

This study investigates the implications of this dark decay channel on neutron star properties. This study takes forward the previous findings and constrains \cite{Husain_2022Conseq, Grinstein:2018ptl, BAYM1971225,Motta:2018bil, Fornal:2018eol} the properties of the dark matter using the gravitational wave, which can be very crucial in determining the interior of the neutron star.

%focusing on baryon number conservation. We aim to derive constraints on dark matter properties and identify potential observational signatures. Section 2 details the equation of state (EoS) and Tolman-Oppenheimer-Volkov (TOV) equations [30–33] used for neutron star modeling. Section 3 explores the impact of the dark decay on neutron star characteristics, comparing results with observational constraints. Finally, Section 4 summarizes our findings and conclusions.

\section{Neutron Stars}
Neutron stars are extremely compact objects that exist in the universe and exhibit an extreme range of energy densities. In fact, black holes are the only objects that are more compact than neutron stars. The energy densities at the outer crust are comparatively lower, and this part plays a crucial role in determining the radius of the neutron stars
At the core, neutron stars contain energy densities several times the energy density of normal nuclear matter. The composition of matter at such high energy densities has been a puzzle for physicists. There have been speculations proposed by different physicists suggesting such high energy densities at the core of neutron stars may contain nuclear matter, hyperons, or even strange matter \cite{Lawley:2006ps,Whittenbury:2013wma,Whittenbury:2015ziz,PhysRevD.4.1601,PhysRevD.30.272,Bombaci_2004,PhysRevD.102.083003,doi:10.1143/JPSJ.58.3555,2012_a,doi:10.1063/1.4909561,2016_a,PhysRevC.58.1804,BALBERG1997435,1985ApJ...293..470G,KAPLAN198657,PhysRevLett.79.1603,PhysRevLett.67.2414,Glendenning1997,Haensel2017,1980PhR....61...71S,weber2007neutron,2019_fri,Weber2016,Terazawa:2001gg,Husain_2021,2001_lattimer,2020_latti,2021_l,Cierniak:2021knt,Shahrbaf:2022upc,10.1143/PTP.108.703,2017_xyz,2022_xxyz,Motta:2022nlj}. But it has been widely accepted that mostly 90\% of it is made of neutrons. This neutron-rich environment renders neutron stars ideal astrophysical laboratories for examining the proposed neutron decay into dark matter (DM) scenarios.

The hypothesis that neutrons may  decay into DM suggests that a substantial number of dark matter particles may  accumulate within neutron stars, assuming dark matter is annihilating, potentially leading to observable alterations in their macroscopic properties \cite{Mukhopadhyay_2017,PhysRevD.77.043515,PhysRevD.77.023006,CIARCELLUTI201119,Sandin_2009,Leung:2011zz,Ellis_2018,bell2020nucleon,2021_w,2000NuPhB.564..185M,Blinnikov:1983gh,2018_sa,2019_reddy,2013_red,berryman2022neutron,McKeen:2021jbh,deLavallaz:2010wp,Busoni:2021zoe,Sen:2021wev,Guha:2021njn}. To model the nuclear matter inside neutron stars with precision, the equation of state (EoS) based on the quark-meson coupling (QMC) model has been used \cite{Guichon:1987jp,Guichon:1995ue,Stone:2016qmi,RIKOVSKASTONE2007341}. This model is rooted in a relativistic mean field. This framework, grounded in relativistic mean-field theory, effectively represents nucleon interactions through meson exchange. In this study, is limited to neutron decay into dark matter, hadronic composition to nucleons and strange quark matter only. %This simplification allows us to focus specifically on the effects of neutron-to-dark matter decay within a well-defined nuclear setting, facilitating a more precise analysis of its influence on neutron star properties. Through this approach, I systematically examine how dark matter accumulation affects observable features of neutron stars, offering valuable insights into dark matter's fundamental characteristics.

\section{Equation of State}
To determine the equation of state, the QMC model has been adapted, which has shown promising results in \cite{Motta:2022nlj,Husain_2022Conseq,Husain_2020,Guichon:2018uew}. The quark-meson coupling (QMC) model was first introduced by Guichon and later expanded upon by Guichon, Thomas, and their collaborators. In this framework, nucleons are modeled as clusters of three quarks confined within an MIT bag, emphasizing their internal structure—unlike other models that treat nucleons as point-like particles \cite{DeGrand:1975cf}. The interactions between baryons arise from meson exchange, with the mesons coupling dynamically to the confined quarks. A key feature of this model is the strong scalar mean field, which induces significant modifications in the properties of bound baryons. %The equation of state derived from the QMC model has been found to provide a reliable description of neutron star characteristics. The details of QMC model can be found in \cite{}.
As proposed by Fornal and Grinstein \cite{Fornal:2020gto,Grinstein:2018ptl}, if neutron stars decay into dark matter. Dark matter must be taken into account when formulating an equation of state. At ($\beta$)-equilibrium, the chemical equilibrium equations governing the composition of matter are \cite{Motta:2018bil,Motta:2019tjc,Husain_2021}:

\begin{equation}
    \mu_n = \mu_\chi,\quad \mu_n = \mu_p + \mu_e ,\quad \mu_\mu = \mu_e
    %μn = μχ + mφ μn = μp + μe μμ = μe np = ne + nμ
\end{equation}
The energy density of the system, including the dark matter, is given by:
\begin{equation}
\begin{aligned} 
    \epsilon_H = \frac{1}{2}m^2_\sigma\sigma^2 + \frac{1}{2}m^2_\omega\omega^2 + 
    \frac{1}{2}m^2_\rho\rho^2 + \frac{1}{\pi^2} \int^{k_F^n}_0 k^2 \sqrt{k^2 + M_N^*(\sigma)^2} dk + \\
    \frac{1}{\pi^2} \int^{k_F^p}_0 k^2 \sqrt{k^2 + M_N^*(\sigma)^2} dk +  \frac{1}{\pi^2}\int^{k_F^e}_0 k^2\sqrt{k^2 + m_e^2} dk +\\ \frac{1}{\pi^2} \int^{k_F^\mu}_0 k^2 \sqrt{k^2 + m_\mu^2} dk + 
    \frac{1}{\pi^2}\int^{k_F^\chi}_0 k^2\sqrt{k^2 + m_\chi^2} dk 
   %  +\tfrac{1}{2}G n_\chi^2 . 
\end{aligned}
\end{equation}
where $m_i$ is the mass of the corresponding particle (subscripted), and the details of the last term are given in section ~\ref{lag}. The nucleon effective mass, $M_N^*(\sigma)$, given by \cite{RIKOVSKASTONE2007341,Husain_2020}
\begin{equation}
M_N^*(\sigma) = M_N -  g_\sigma \sigma + 
\frac{d\tilde(g_\sigma \sigma)^2}{2} \, . 
\label{eq2.1}
\end{equation}
Here $g_\sigma$ is the $\sigma$-nucleon coupling constant in free space calculated in the model and $d$ is the scalar polarizability parameter of the nucleon in the QMC model, which accounts for quadratic corrections to the scalar field coupling \cite{RIKOVSKASTONE2007341}.
The pressure ($P$) is calculated using
\begin{equation}
P = \int_{i} \mu_i n_i - \varepsilon,
\label{eq:pressure}
\end{equation}

where $i = n,p,e,\mu,\chi$. The nucleons-only EoS is taken from \cite{RIKOVSKASTONE2007341} and strange matter EoS is drive based on MIT bag model, and the EoS used is given by
\begin{equation}
P = \frac{1}{3}(\varepsilon - 4B),
\label{eq:bag_model}
\end{equation}
where $\epsilon$ the energy density of the strange matter and $B$ is the bag constant, taken to be  57$~\text{MeV/fm}^3$ \cite{Urbanec_2013}. 
%\appendix
\section{Structural equations}
Neutron stars are dense astrophysical objects shaped by intense nuclear and gravitational forces. The potential presence of dark matter within these stars may influence their structural characteristics, resulting in changes to their equation of state (EoS). This study examines methods for integrating dark neutron components into the conventional neutron star EoS and solving the Tolman-Oppenheimer-Volkoff (TOV) \cite{Tolman169,PhysRev.55.374} equations to assess their impact.
\subsection*{TOV equation}
%\section{Tolman-Oppenheimer-Volkoff Equations}
%To model the structure of a neutron star including dark matter effects, we solve the TOV equations:
\begin{equation}
\frac{dP}{dr} = -\frac{(\epsilon + P)}{r}\frac{d\Phi}{dr}
\end{equation}

\begin{equation}
\frac{d\Phi}{dr} = \frac{M + 4\pi r^3 P}{r(r - 2M)}
\end{equation}

\begin{equation}
\frac{dM}{dr} = 4\pi r^2 \epsilon
\end{equation}
where $P$ is the pressure, $\epsilon$ is the energy ensity, $\Phi$ is the gravitational potential, $M$ is the mass, and $r$ is the radial coordinate.

\subsection*{Perturbation Equations for f-mode Oscillations
}
The Lindblom–Detweiler formalism for non-radial oscillations \cite{1983ApJS...53...73L} has been adopted in this study. Here $\nu(r)$ and $\lambda(r)$ are the static metric functions in the line element $ds^2=-e^{\nu}dt^2+e^{\lambda}dr^2+r^2 d\Omega^2$, and $(X,W,V)$ are perturbation variables defined as in \cite{1985ApJ...292...12D}. This ensures dimensional consistency and reproducibility.

The f-mode oscillations are governed by two coupled differential equations \cite{Jaiswal_2021,Celato_2025,Guha_Roy_2024}
\begin{equation}
\frac{dX}{dr} = -\frac{l+1}{r} X + \frac{e^{\lambda/2}}{r} \left[ \omega^2 r e^{-\nu/2} V + \frac{d\nu}{dr} W \right],
\end{equation}
\begin{equation}
\frac{dW}{dr} = -\frac{d\nu}{dr} W - \frac{l(l+1)}{r} V + e^{\lambda/2} X,
\end{equation}

%\begin{equation}
%\frac{dW}{dr} = \frac{d\epsilon}{dP} \left[ \omega^2 r^2 e^\lambda - 2\Phi V + \frac{d\Phi}{dr} W \right] - l(l+1) e^\lambda V
%\end{equation}

%\begin{equation}
%\frac{dV}{dr} = 2 \frac{d\Phi}{dr} V - e^\lambda \frac{W}{r^2}
%\end{equation}
where $W(r)$ represents the radial displacement function, $V(r)$ is the metric perturbation function, $\omega$ is the angular oscillation frequency, and $l$ is the spherical harmonic index of the perturbation.

\subsection*{Boundary Conditions and Numerical Integration}
To solve these equations, boundary conditions are applied:
\begin{itemize}
\item At the center:
\begin{equation}
W(r) = A r^{l+1}, \quad V(r) = -A r^l / l
\end{equation}
\item At the surface:
\begin{equation}
\omega^2 R^2 e^\lambda - 2\Phi V + \frac{d\Phi}{dr} W = 0
\end{equation}
\end{itemize}
The equations are integrated numerically. The eigenvalues of the system correspond to the f-mode frequencies.
%\subsection{Numerical procedure}To compute the f-mode eigenfrequencies, we solved the perturbation equations (Eqs.~\ref{eq:pert1}–\ref{eq:pert4}) as a boundary value problem using a standard shooting method. Starting from the center, the regular series expansion (Eq.~\ref{eq:series}) was used to initialize the variables, and the system was integrated outward with adaptive Runge–Kutta. The trial frequency $\omega$ was adjusted iteratively until the surface boundary condition $X(R)=0$ was satisfied within a relative tolerance of $10^{-6}$. Convergence was verified by repeating the integration with smaller step sizes and ensuring eigenfrequencies were stable at the $\lesssim 0.1\%$ level across resolutions and for different equations of state.

%For a two-fluid system with baryonic and dark components, the coupled TOV equations take the form:
%\begin{align}
%    \frac{dP_b}{dr} &= -\frac{G (\rho_b + P_b/c^2)(M + 4\pi r^3 (P_b + P_d)/c^2)}{r^2(1 - 2GM/rc^2)}, \\
%    \frac{dP_d}{dr} &= -\frac{G (\rho_d + P_d/c^2)(M + 4\pi r^3 (P_b + P_d)/c^2)}{r^2(1 - 2GM/rc^2)}.
%\end{align}
% 
%\section{Numerical Implementation}
%The system of equations can be solved using the Runge-Kutta method or a shooting method with boundary conditions:
%\begin{itemize}
%    \item $P(r=0) = P_c$ (central pressure)
%    \item $M(r=0) = 0$ (mass at the center)
%    \item $P(R) = 0$ for the neutron star radius $R$
%\end{itemize}

\section{Relationship between energy density and pressure}
\subsection{Effective Lagrangian for neutron to dark matter decay}\label{lag}

To model the possibility of neutron decay into dark matter inside neutron stars, 
The nucleonic sector (described here using the QMC model) has been extended by including 
a dark fermion $\chi$ carrying baryon number. In order to reproduce the repulsive self-interaction assumed in the EoS analysis, a dark vector mediator $V_\mu$ has been introduced, 
analogous to the $\omega$ meson in relativistic mean-field models. The effective 
Lagrangian reads
\begin{equation}
\mathcal{L} \;=\; \mathcal{L}_{\rm QMC}[n,p,\sigma,\omega_\mu,\rho_\mu]
+ \bar{\chi}\,(i\slashed{\partial}-m_\chi)\,\chi
-\frac{1}{4} V_{\mu\nu} V^{\mu\nu}
+ \frac{1}{2} m_V^2 V_\mu V^\mu
- g_\chi\,\bar{\chi}\gamma^\mu \chi V_\mu
+ \varepsilon\,(\bar{n}\chi + \bar{\chi}n),
\end{equation}
where $V_{\mu\nu}=\partial_\mu V_\nu-\partial_\nu V_\mu$. 
The last term represents an effective in-medium mixing between neutrons and dark fermions, 
which allows for neutron conversion into dark matter when $\mu_n > \mu_\chi$. 
Although the effective in-medium operator $\varepsilon \bar{n}\chi + \text{h.c.}$ mixes neutrons with dark fermions, baryon number is conserved if the dark fermion $\chi$ carries baryon number $B(\chi)=1$. In this case, the conversion $n \leftrightarrow \chi$ redistributes baryon number between visible and dark sectors but does not violate overall conservation. This treatment follows the effective approach proposed in \cite{Fornal:2018eol,Fornal:2020gto,Grinstein:2018ptl}.

%Baryon number is conserved if $\chi$ carries baryon number $=1$.

\begin{figure}[h!]
    \centering
    \includegraphics[width=1\linewidth]{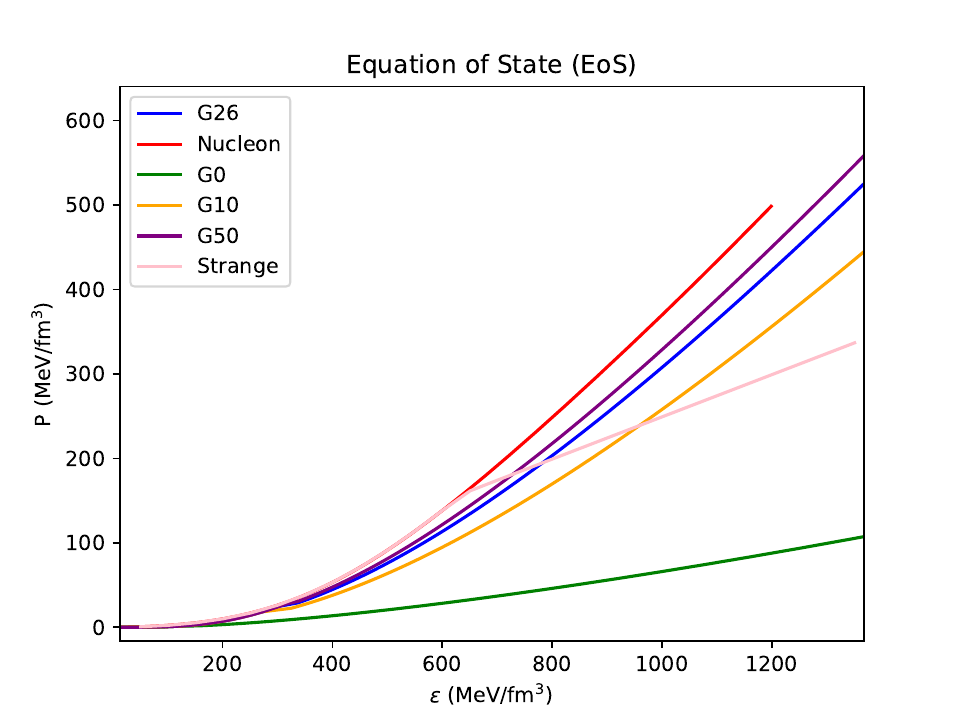}
    \caption{Equations of state (pressure vs. energy density) for nucleonic matter (QMC), strange matter (MIT bag model), and neutron decay into dark matter for different DM–DM self-interaction strengths $G$ (fm$^2$). Higher $G$ corresponds to stronger repulsive self-interactions.}
    \label{fig:eos}
\end{figure}

At scales below $m_V$, integrating out the heavy vector yields a four-fermion operator,
\begin{equation}
\mathcal{L}_{\rm eff} \supset 
\frac{g_\chi^2}{2m_V^2}\,(\bar{\chi}\gamma_\mu \chi)(\bar{\chi}\gamma^\mu \chi) 
\equiv \frac{G}{2}\,(\bar{\chi}\gamma_\mu \chi)^2,
\label{eq:Geff}
\end{equation}
where the effective coupling is defined as $G=g_\chi^2/m_V^2$. 
In mean-field approximation, this produces a repulsive contribution 
to the energy density and pressure, 
$\Delta\varepsilon=\tfrac{1}{2}G n_\chi^2$, $\Delta P=\tfrac{1}{2}G n_\chi^2$,
which stiffens the equation of state as $G$ increases \cite{Motta:2018rxp,Motta:2018bil,Husain_2022Conseq}. 
This mechanism directly explains why the dark-matter–induced EoS approaches 
the nucleon-only EoS for large $G$, as shown in Figure (\ref{fig:eos}), which presents the equations of state (EoS) for nucleonic matter, strange matter, and scenarios involving neutron decay into dark matter under varying dark matter self-interactions. For the nucleon-only case, the Quark-Meson Coupling (QMC) model is employed. The EoS describing neutron decay into dark matter is shown for several values of the dark matter–dark matter (DM–DM) self-repulsion strength. For comparison, the strange matter EoS is also included.

The results indicate that the nucleon-only EoS is the stiffest among the considered models. As the DM–DM self-repulsion increases, the corresponding EoS for neutron decay into dark matter becomes progressively stiffer and approaches the nucleon-only case, which is in agreement with the findings of \cite{Husain_2022Conseq,Motta:2018rxp,BAYM1971225}. This behavior arises because stronger DM–DM self-interaction raises the energy cost for neutron decay into dark matter, thereby reducing the overall dark matter population within the system \cite{Husain_2022Conseq}. Consequently, larger self-interaction strengths, $G$, result in fewer dark matter particles and stiffer equation of state \cite{Husain_2022Conseq}.

%In this framework, the dark matter self-interaction is modeled analogously to the neutron–$\omega$ coupling, similar to previous studies \cite{Motta:2018rxp,Motta:2018bil,Husain_2022Conseq}, represented by $G$. It is defined as:
%\begin{equation}
%     G = (\frac{g_i}{m_i})^2 
%\end{equation}
%where $g_i$ is the coupling constant, and $m_i$ is the mass of the interchange particle. Various values of $G$ are selected in this study, where the number next to $G$ represents the strength of the DM-DM particle interaction. A higher $G$ value indicates stronger DM-DM self-interaction, influencing the overall dynamics and structural properties of the system.

%The strange matter EoS is based on the MIT bag model and produces taken from \ref{}. 
The strange matter EoS provides a poor description of nuclear matter at low energy densities. As shown in Fig. (\ref{fig:eos}), it coincides with the nucleon-only EoS up to approximately 500 MeV/fm$^3$ \cite{RIKOVSKASTONE2007341,Husain_2020,Husain_2022Conseq}. Beyond this density, strange matter begins to manifest, and the corresponding EoS deviates from both the nucleon-only and the neutron-decay EoSs. In this regime, the strange matter EoS is comparatively softer than the other two models. The deviation of the strange matter EoS at densities above $\sim500$ MeV/fm$^3$ arises from the bag constant assumption in the MIT bag model, which lowers the pressure support compared to nucleonic models. As a result, the sound speed is reduced, leading to longer damping times and lower f-mode frequencies. This implies that strange stars may emit weaker gravitational waves, with potentially distinguishable signals compared to nucleonic stars of the same mass.

\begin{figure}[h!]
    \centering
    \includegraphics[width=1\linewidth]{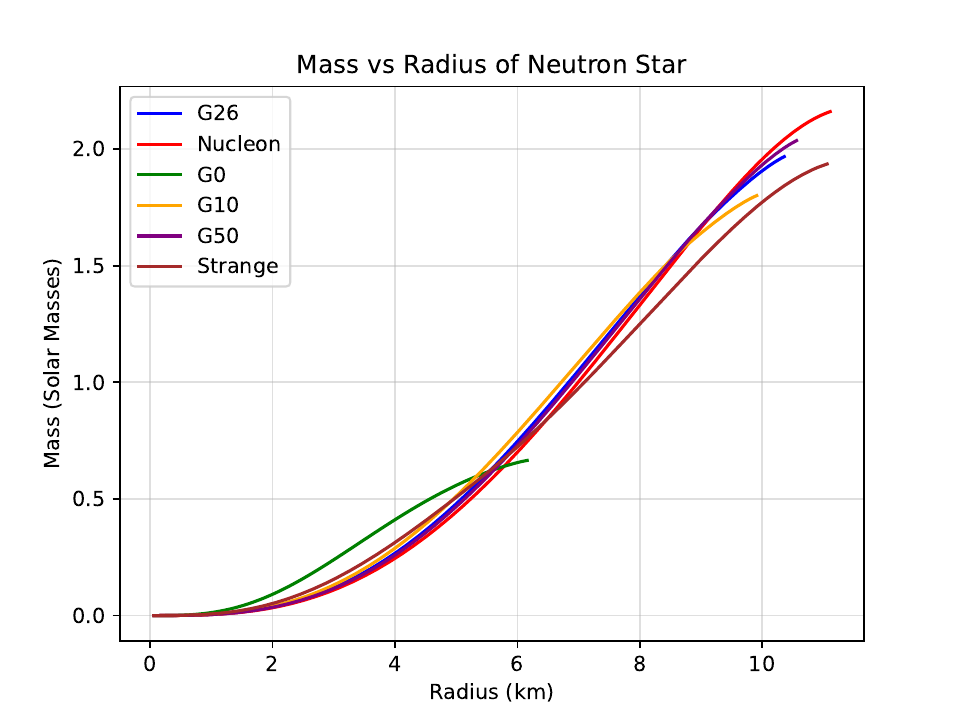}
    \caption{Mass–radius relations for neutron stars with different equations of state (EoSs). Shown are nucleon-only (QMC), strange matter (MIT bag model), and dark-matter–admixed stars with varying self-interaction strengths $G$.  The radius of light neutron stars is smaller because the EoS is not optimized in the lower energy density region (outer crust region) \cite{Husain_2020}.}
    \label{fig:MvsR}
\end{figure}

Fig. (\ref{fig:MvsR}) represents the mass and radius of neutron stars based on different EoSs. Ref. \cite{Ozel:2016oaf,2010Natur.467.1081D,2013Sci...340..448A} report the observation of neutron stars of a mass of exceeding 2 M$_\odot$ \cite{2019ApJ...887L..24M,
2019ApJ...887L..21R,2018PhRvL.121p1101A,2021ApJ...918L..28M}, indicating that any viable model must support a maximum neutron star mass above this threshold 2 M$_\odot$. The nucleon-only EoS produces a neutron star of maximum mass greater than 2 M$_\odot$, the strange matter EoS produces a neutron star of maximum mass close to 2 M$_\odot$, while different EoSs based on different DM-DM self-interactions produce a neutron star of less than 2 M$_\odot$. The non-self-interactive ($G$ = 0 fm$^2$) DM EoS produces a neutron star of mass close to 0.7 M$_\odot$ which is well below the acceptable value and alignes with \cite{Motta:2018rxp,Grinstein:2018ptl,BAYM1971225}. When the DM-DM self-interaction $G = 26\,\mathrm{fm}^2$ \cite{Husain_2022Conseq} the neutron decay into dark matter EoS produces a neutron star of 2 M$_\odot$. Therefore,  acceptable self-interaction strength is $G >$ 26 fm$^2$.
The failure of the $G=0$ fm$^2$ case to support neutron stars above $\sim0.7 M_\odot$ highlights that non-self-interacting dark fermions dilute pressure support too severely. Physically, the DM particles form a diffuse distribution, decreasing compactness and oscillation frequencies. 
%This explains why the universal relations break down in this limit, reinforcing the need for DM self-repulsion in any viable neutron-decay model.
\section{Universalities}
The universalities refer to empirical relations that link the fundamental oscillation modes of compact objects, such as neutron stars and dark stars, to their macroscopic properties like mass, radius, and compactness. These universal relations are valuable because they provide a model-independent way to infer stellar characteristics from gravitational wave observations \cite{dey2024fmodeoscillationsdarkmatter,Celato_2025}. Traditional neutron stars follow well-established scaling laws for f-mode frequencies and damping times, but recent studies show that dark matter admixed compact stars deviate from these relations, requiring new frameworks to describe their oscillations \cite{PhysRevD.110.063025,Jyothilakshmi2025}. By leveraging f-mode gravitational wave signals, astrophysicists aim to refine equations of state and probe exotic forms of matter beyond the Standard Model. The detailed analysis on tidal deformability of neutron decay mode is given in \cite{Husain_2022Conseq}.
\subsection{Damping time}
The damping time $\tau$ of f-mode oscillations is related to the balance between the total mode energy $E$ and the rate of energy loss via gravitational-wave emission $\dot{E}{\rm_{GW}}$,
\begin{equation}
\tau = \frac{2E}{\dot{E}_{GW}}.
\end{equation}
The oscillation energy for a mode with angular frequency $\omega$ is given by
\begin{equation}
E = \frac{1}{2} \omega^2 \int \rho(r) |\xi(r)|^2 d^3r,
\end{equation}
where $\xi(r)$ is the Lagrangian displacement vector and $\rho(r)$ the local energy density. The gravitational-wave luminosity for the dominant quadrupole mode is
\begin{equation}
\dot{E}{\rm _{GW}} = - \frac{32 G}{5c^5} \omega^6 \left| \int \rho(r) r^2 Y{2m}(\theta,\phi) dr d\Omega \right|^2.
\end{equation}
From these relations, the universal scaling $\tau \sim R^4/M^3$ emerges \cite{Andersson1998}, which has been adopted and extended in this study. %The full perturbation equations (Eqs. 5.4–5.5) are solved numerically using shooting methods with boundary conditions (Eqs. 5.6–5.7), and the eigenfrequencies $\omega$ extracted correspond to the f-mode oscillations.
\par
\begin{figure}[h!]
    \centering
    \includegraphics[width=1\linewidth]{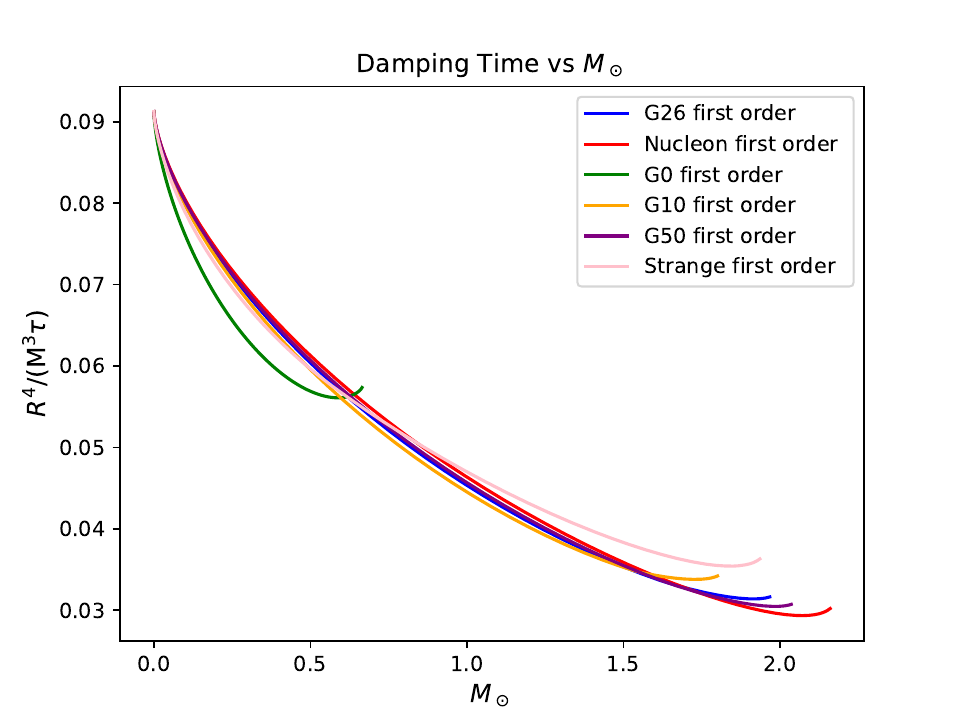}
    \caption{The normalised damping time of the f-mode as a function of stellar mass. Nucleon-only stars exhibit the shortest damping times, while strange-matter stars show longer-lived oscillations. }
    \label{Fig2}
\end{figure}

Reference \cite{10.1046/j.1365-8711.1998.01840.x} introduced a first-order universal relation for the damping time ($\tau$) of f-mode oscillations, derived from the quadrupole radiation formula. The relation, shown in Figure (\ref{Fig2}), is expressed as
\begin{equation}
\tau \sim \frac{R^4}{M^3}.
\label{damp}
\end{equation}
Here $\tau$ reflects the ratio of oscillation energy to gravitational-wave luminosity, implying that more massive stars radiate more efficiently and exhibit shorter damping times. Lioutas and Stergioulas \cite{Lioutas2018} later extended this relation with higher-order corrections, confirming the monotonic decrease of $\tau$ with mass. In comparison, strange matter stars exhibit longer damping times than both nucleon-only stars and neutron stars with dark matter cores.

\subsection{Mass-scaled angular frequency}
\begin{figure}[h!]
    \centering
    \includegraphics[width=1\linewidth]{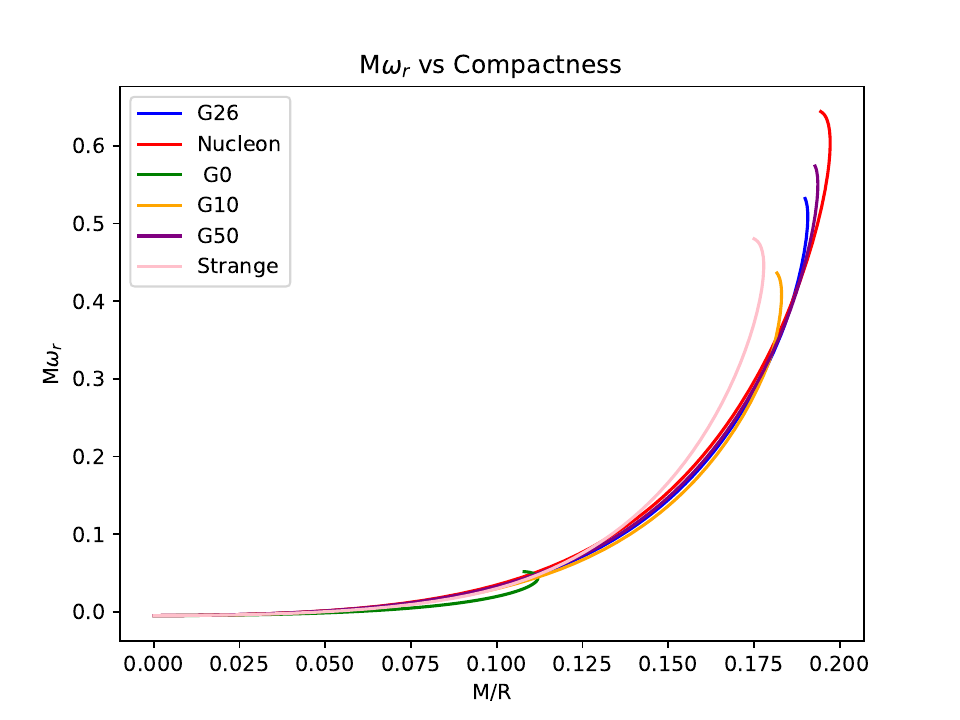}
    \caption{The relationship between mass-scaled angular frequency and compactness, where mass is given in solar masses and radius in given in kilometers, demonstrating a universal trend, is plotted for selected dark matter equations of state (EoSs). This analysis includes various DM-DM self-interaction strengths.}
    \label{Mwr1}
\end{figure}
\begin{figure}[h!]
    \centering
    \includegraphics[width=1\linewidth]{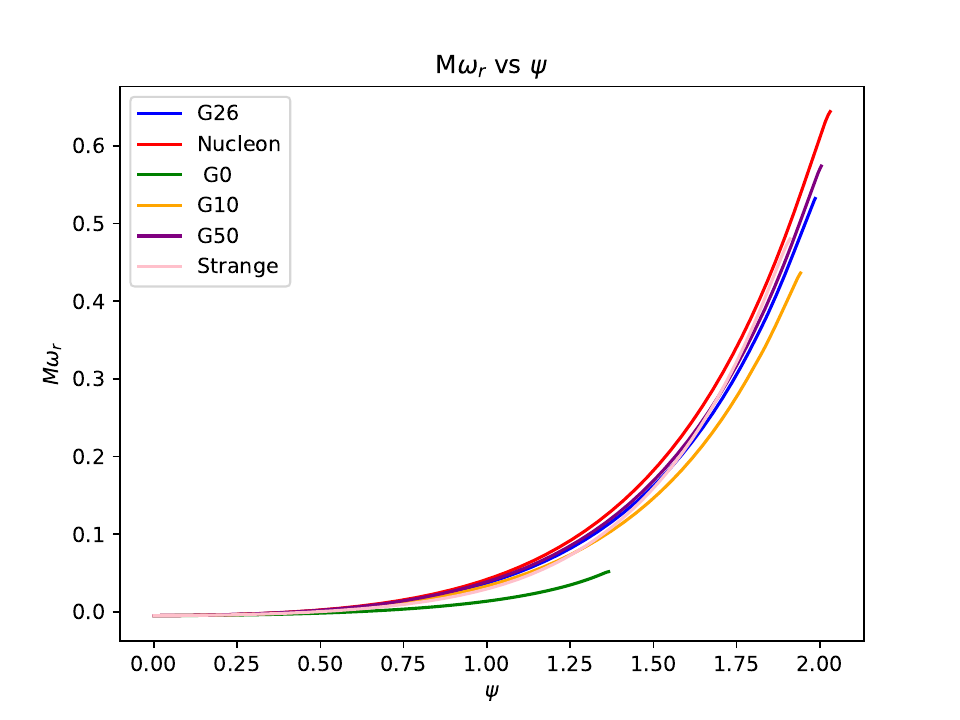}
    \caption{The relationship between mass-scaled angular frequency and effective compactness is plotted for selected dark matter EoSs, including various DM-DM self-interaction strengths. }
    \label{Mwr2}
\end{figure}
Figures (\ref{Mwr1}) and (\ref{Mwr2}) illustrate the relation between the mass-scaled angular frequency, where $\omega_r = 2\pi f$, and two structural parameters of neutron stars, namely, compactness ($M/R$) and $\psi=\sqrt{M^3/I}$ \cite{Celato_2025}. These diagnostics are particularly valuable since they encode information about the distribution of mass and the role of internal composition in determining stellar oscillation properties. The parameter $\sqrt{M^3/I}$, in particular, is closely tied to the efficiency of gravitational-wave (GW) emission, as it relates the global stellar mass to the moment of inertia, thereby providing insight into how matter is distributed inside the star.\par
As Figure (\ref{Mwr1}) shows, when $G>$ 26 fm$^2$, the nucleon-only and self-interacting DM EoSs follow a universal scaling curve, independent of the specific interaction strength. This suggests that compactness is a reliable predictor of oscillation properties, robust against moderate variations in microphysics. Strange matter and non-self-interacting DM deviate significantly. The strange matter deviation reflects the reduced stiffness of the EoS, while the DM $G=0$ case is destabilized by the large DM population at the core, altering the star’s effective compactness. These deviations may serve as diagnostic signatures of exotic compositions.

Figure (\ref{Mwr2}) further reinforces the presence of a quasi-universal relation across different microscopic models when the angular frequency is expressed in terms of $\psi=\sqrt{M^3/I}$. Remarkably, even when exotic compositions are considered, including strange matter and neutron-decay into DM with strong self-interaction ($G>26$ fm$^2$), the data points collapse onto a nearly single curve. This indicates that $\psi$ effectively captures the essential structural scaling of neutron stars, smoothing out differences in composition. Deviations are observed only in the dark matter without self-repulsion case, but the overall correlation persists. This reinforces the robustness of universal relations in gravitational-wave asteroseismology.

\section{F-mode frequency}
\begin{figure}[h!]
    \centering
    \includegraphics[width=1\linewidth]{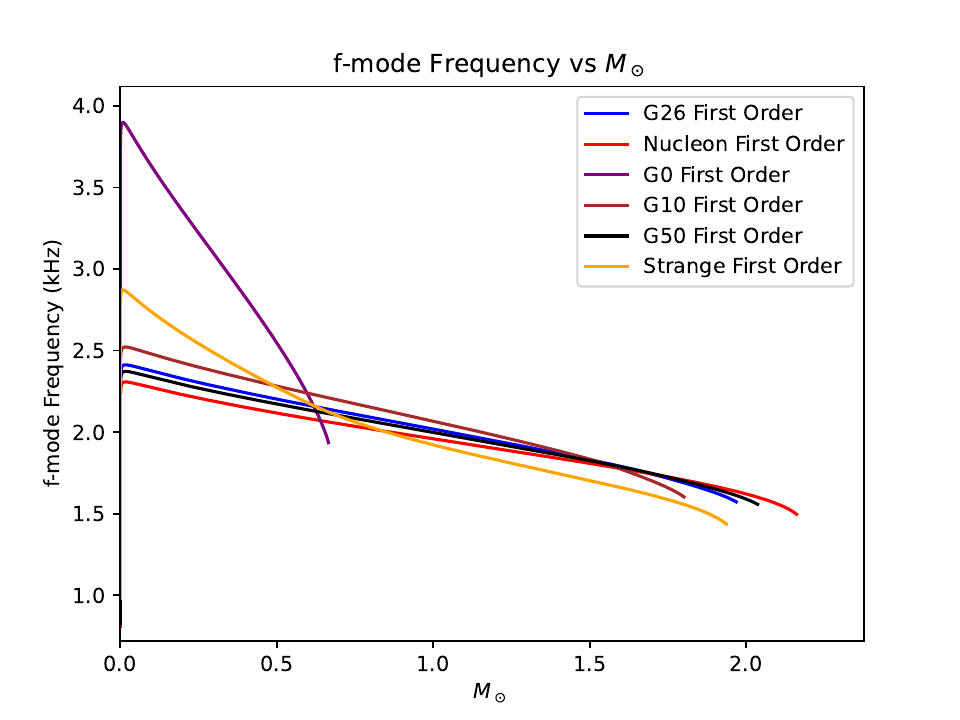}
    \caption{The f-mode frequencies are calculated for various equations of state (EoSs), taking into account different dark matter (DM) self-interaction strengths, agreeing to Ref. \cite{Lau2010,Lioutas2018}.}
    \label{f_mode_1}
\end{figure}

The frequency of an f-mode is highly sensitive to the EoS of the ultra-dense matter within the neutron star \cite{Jaiswal_2021,Celato_2025,Guha_Roy_2024}. %Different theoretical EoS models predict different f-mode frequencies for a given neutron star mass. 
A stiffer EoS generally predicts higher f-mode frequencies. Conversely, a softer EoS predicts lower f-mode frequencies. The acceptable values of f-mode frequency for compact stars, such as neutron stars, typically range between 1 and 1–3 kHz. These frequencies are significant because they fall within the detection range of advanced gravitational wave detectors like the Einstein Telescope and Cosmic Explorer \cite{Punturo_2010}. Figure (\ref{f_mode_1}) illustrates that the f-mode frequency of the selected models adheres to the universal relation, demonstrating consistency across different astrophysical scenarios.

Notably, both frequency ranges fall within the detection capabilities of the Einstein Telescope and Cosmic Explorer, highlighting their potential for observational verification.
Furthermore, for neutron decay into dark matter, variations in dark matter self-interaction strength do not disrupt the universal trend. Regardless of the DM-DM self-interaction parameter, the f-mode frequency remains well within the expected range, reinforcing the robustness of the underlying physical principles governing these oscillations. This consistency suggests that such models could be effectively probed through next-generation gravitational wave observatories.\par
%Our results are consistent with earlier works exploring neutron-to-dark matter decay in neutron stars. Motta et al. \cite{Motta:2018bil} and Grinstein et al. \cite{Grinstein:2018ptl} found that non-interacting dark fermions destabilize neutron stars, reducing their maximum mass below the $2M_\odot$ threshold, which we confirm in our analysis. Husain et al. \cite{Husain_2022Conseq} similarly emphasized the necessity of DM self-repulsion, consistent with our finding that $G>26$ fm$^2$ restores compatibility with observational constraints.\par
For f-mode oscillations, this study extends the universal relations established by Andersson and Kokkotas \cite{Andersson1998} and refined by Lioutas and Stergioulas \cite{Lioutas2018}, showing that these relations persist even in the presence of exotic matter. However, it has been demonstrated that strange matter and non-self-interacting DM produce systematic deviations, suggesting that precise future detections could be used to discriminate between EoSs.

\begin{figure}[h!]
    \centering
    \includegraphics[width=1\linewidth]{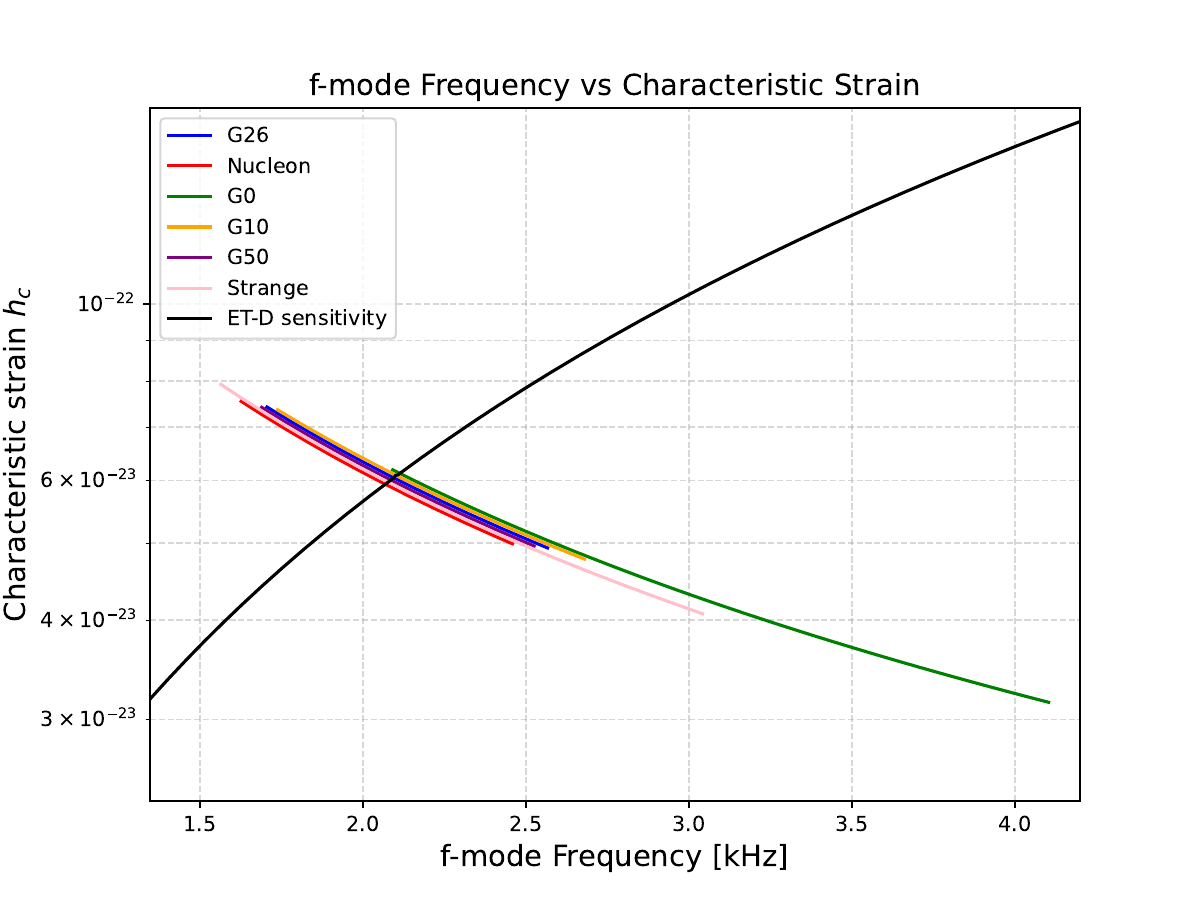}
    \caption{Characteristic strain $h_c$ of f-mode oscillations from neutron stars as a function of frequency, compared with the sensitivity curve of the Einstein Telescope (ET-D) \cite{ETsensitivities}. Results are shown for several equations of state (EoS), assuming a total gravitational-wave energy release of E = 10$^{52}$ erg, damping time = 0.1 sec and a source distance of D = 25 Mpc.}
    \label{ET sensitivity}
\end{figure}

Figure \ref{ET sensitivity} presents the predicted characteristic strain of f-mode oscillations for the considered EoS models, compared with the ET-D sensitivity curve \cite{Hild_2011,ETsensitivities}. The strain values have been calculated assuming a fiducial gravitational-wave energy release of E = 10$^{52}$ erg, a source distance of D = 25Mpc, and a damping time of $\tau$ = 0.1s, consistent with estimates in \cite{zheng2025fmodeoscillationsprotoneutronstars,Li_2025}. Under these benchmark conditions, the non–self-interacting dark matter EoS (G0) yields strain amplitudes that remain below the ET sensitivity threshold across the relevant frequency range, implying that such signals would not be detectable. By contrast, the self-interacting dark matter and strange matter EoSs predict higher characteristic strains that exceed the ET sensitivity curve for frequencies below  2.1 kHz. This demonstrates that, if sources within 25 Mpc deposit of order 
10$^{52}$ erg into f-mode oscillations, the corresponding signals could in principle be detected by ET. These results emphasize that f-mode detectability is highly sensitive to the underlying neutron-star EoS, and they suggest that next-generation detectors such as ET can provide meaningful constraints on dark matter interactions through neutron-star asteroseismology. 
%Figure (\ref{ET sensitivity}) illustrates the predicted characteristic strain of f-mode oscillations relative to the ET-D sensitivity curve \cite{Hild_2011,ETsensitivities}. For the non–self-interacting dark matter EoS (G0), the strain amplitudes lie consistently below the ET sensitivity threshold, implying that such signals would not be detectable by ET under the assumed conditions. In contrast, the other EoS models predict higher characteristic strains, which exceed the ET sensitivity curve at frequencies below approximately  2.1 kHz. This result indicates that f-mode gravitational-wave signals from these models would, in principle, fall within the detectable range of ET, provided the source lies within D = 25 Mpc and the energy release is of order 10$^{52}$ erg with GW damping time = 0.1 sec, which is consistent with the References \cite{zheng2025fmodeoscillationsprotoneutronstars, Li_2025}. These findings highlight the sensitivity of f-mode detectability to the underlying neutron star equation of state and suggest that future third-generation detectors such as ET could place meaningful constraints on dark matter interactions through neutron star asteroseismology.

\begin{figure}[h!]
    \centering
    \includegraphics[width=1\linewidth]{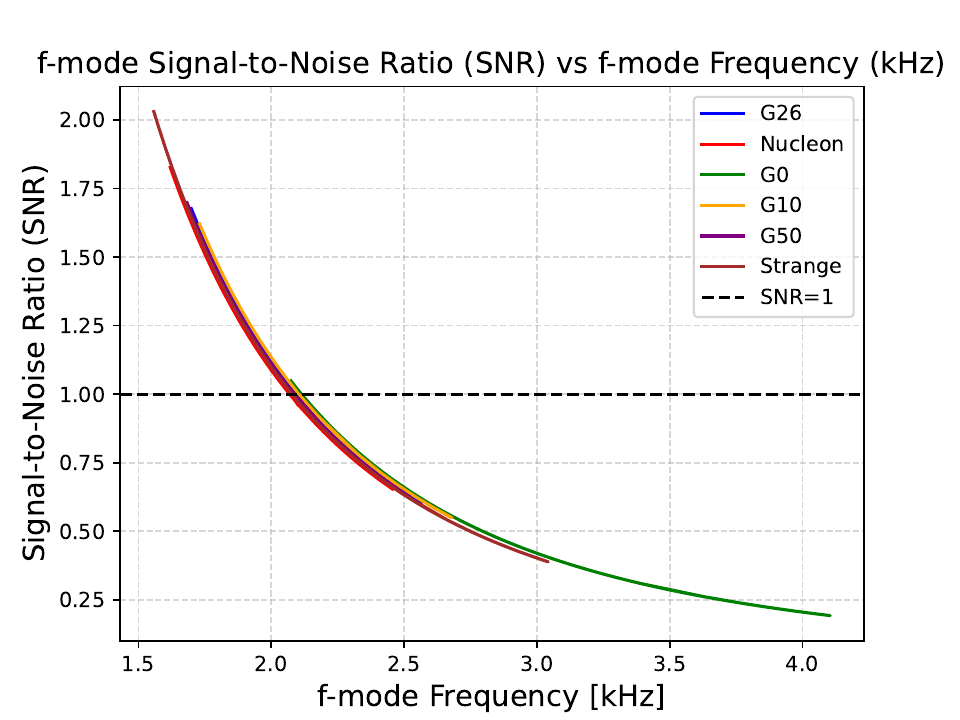}
    \caption{Signal-to-noise ratio (SNR) of f-mode gravitational-wave signals as a function of frequency, computed for different equations of state (EoS) at a source distance of D = 25 Mpc, damping time = 0.1 sec and energy release E = 10$^{52}$ erg.}
    \label{SNR}
\end{figure}

Figure (\ref{SNR}) shows the variation of SNR with f-mode frequency for the different EoS models considered. It has been found that the SNR exceeds unity for frequencies below approximately 2.1 kHz. This implies that, within this frequency range, f-mode oscillations could be detectable by the Einstein Telescope, provided the assumed energy budget and source distance. At higher frequencies, the SNR falls below unity, indicating that detection becomes unlikely due to the reduced strain amplitude and the rising noise floor of the detector. Notably, the non–self-interacting DM model (G0) yields systematically lower SNRs, while the interacting dark matter and strange matter EoS predict higher SNR values, suggesting more favorable detectability. These results emphasize that the detection prospects of f-mode signals strongly depend on both the EoS and the frequency band of emission, with ET’s sensitivity window favoring signals in the 2.1 kHz regime.

%\section{Discussion}

The detectability of f-mode gravitational-wave signals is highly sensitive to both the neutron star equation of state (EoS) and the frequency band in which the mode energy is emitted. %Figure (\ref{ET sensitivity}) illustrates the characteristic strain of f-mode oscillations for different EoS models compared against the ET-D sensitivity curve. For the non–self-interacting dark matter case (G0), the predicted strains remain consistently below the ET sensitivity threshold, indicating that such signals would not be observable under the assumed conditions of $E = 10^{52}$ erg and $D = 25$ Mpc. In contrast, the nucleonic, strange matter, and self-interacting DM EoS models yield higher strain values, exceeding the ET sensitivity at frequencies below $\sim 2.1$ kHz. This frequency dependence is consistent with the expected response of third-generation detectors, which achieve their optimal sensitivity in the kilohertz regime.

%The corresponding signal-to-noise ratios (SNR), shown in Figure \ref{SNR}, reinforce this conclusion. For all interacting dark matter and strange matter models, the SNR exceeds unity for $f \lesssim 2.1$ kHz, suggesting that these signals would be detectable by ET under favorable astrophysical conditions. At higher frequencies, however, the SNR drops below unity due to both the reduced strain amplitude and the rising detector noise floor. The G0 model once again stands out as systematically weaker across the frequency spectrum, underscoring the impact of dark matter self-interactions on detectability.

The physics behind these trends can be understood as follows. Softer equations of state, such as those with exotic matter components, generally produce lower f-mode frequencies, shifting the emission closer to the frequency band where ET is most sensitive. Moreover, longer damping times associated with strange matter enhance the overall energy radiated in the mode, further increasing the characteristic strain. Conversely, the absence of DM self-repulsion (G0) leads to reduced pressure support, suppressing the excitation energy available to the mode and lowering the strain amplitude. Thus, the inclusion of dark matter with repulsive self-interactions not only restores astrophysical stability but also improves the prospects for gravitational-wave detection.

From an observational standpoint, these results indicate that ET has the potential to place meaningful constraints on the nature of dark matter interactions and the composition of neutron stars. A confirmed detection of f-mode oscillations in the $\sim$1–2 kHz range would favor scenarios involving exotic matter or self-interacting dark matter, whereas a systematic absence of such signals could be used to rule out or constrain these models. Future work should extend this analysis to population studies, varying source distances, masses, and energy release efficiencies, in order to more comprehensively assess the detectability of f-modes in realistic astrophysical environments.

\section{Conclusion}

In this work, the f-mode oscillation properties have been explored of neutron stars containing nucleonic matter, strange matter, and dark matter produced through neutron decay, with particular attention to the role of dark matter self-interactions. By computing f-mode frequencies, damping times, characteristic strains, and signal-to-noise ratios, the detectability of these oscillations with the Einstein Telescope (ET) is assessed.
\par The nucleonic EoS remains the stiffest, producing systematically higher f-mode frequencies, while the inclusion of dark matter generally softens the stellar structure. Strong DM–DM self-repulsion ($G \gtrsim 26$ fm$^2$) counteracts this softening and restores compatibility with astrophysical mass constraints. Strange matter predicts longer damping times and lower frequencies, shifting the signal toward the ET’s most sensitive frequency band.
\par For an assumed gravitational-wave energy release of $E = 10^{52}$ erg, damping time = 0.1 sec and a source distance of $D = 25$ Mpc, the characteristic strain and SNR of self-interacting dark matter and strange matter models exceed the ET sensitivity threshold below $\sim 2.1$ kHz. In contrast, the non–self-interacting DM case (G0) lies below the sensitivity curve, suggesting that such signals would be undetectable.
\par These results indicate that third-generation gravitational-wave observatories could realistically probe exotic matter scenarios through f-mode asteroseismology. A confirmed detection of f-mode oscillations in the $\sim$1–2 kHz range would provide strong evidence for exotic matter in neutron star interiors, whereas a lack of such signals could place meaningful constraints on dark matter self-interaction strengths and the viability of strange matter.
\par This study demonstrates that f-mode oscillations represent a promising observational probe of neutron star interiors and exotic matter physics. The persistence of universal relations across different EoS models highlights their robustness, while the deviations introduced by strange matter and non-self-interacting dark matter suggest possible observational discriminants. Future detections of f-mode gravitational waves by instruments such as ET and Cosmic Explorer could therefore open a new window into the microphysics of dense matter, providing unprecedented insight into the role of dark matter and exotic phases in compact stars. This analysis extends previous work by explicitly connecting exotic matter EoS models to gravitational-wave observability, making this the first quantitative detectability study of neutron-decay dark matter through f-mode oscillations. The results indicate that third-generation interferometers such as ET could provide decisive tests of dark matter self-interactions and the presence of exotic matter in compact stars.

\section*{Acknowledgment}
I sincerely thank Professor Anthony Thomas (University of Adelaide) for his invaluable corrections and insightful suggestions, which have significantly enhanced the clarity and accuracy of this work. His expertise and guidance were instrumental in refining our analysis, and I deeply appreciate his suggestion to this study.

\bibliography{main}

\begin{thebibliography}{100}

\bibitem{Moore_1999}
B.~Moore, S.~Ghigna, F.~Governato, G.~Lake, T.~Quinn, J.~Stadel, and P.~Tozzi,
  ``Dark matter substructure within galactic halos,'' {\em The Astrophysical
  Journal}, vol.~524, p.~L19, sep 1999.

\bibitem{Navarro_1997}
J.~F. Navarro, C.~S. Frenk, and S.~D.~M. White, ``A universal density profile
  from hierarchical clustering,'' {\em The Astrophysical Journal}, vol.~490,
  p.~493, dec 1997.

\bibitem{Klypin_1999}
A.~Klypin, A.~V. Kravtsov, O.~Valenzuela, and F.~Prada, ``Where are the missing
  galactic satellites?,'' {\em The Astrophysical Journal}, vol.~522, p.~82, sep
  1999.

\bibitem{10.1093/mnras/264.1.201}
G.~Kauffmann, S.~D.~M. White, and B.~Guiderdoni, ``The formation and evolution
  of galaxies within merging dark matter haloes,'' {\em Monthly Notices of the
  Royal Astronomical Society}, vol.~264, pp.~201--218, 09 1993.

\bibitem{10.1111/j.1745-3933.2011.01074.x}
M.~Boylan-Kolchin, J.~S. Bullock, and M.~Kaplinghat, ``Too big to fail? the
  puzzling darkness of massive milky way subhaloes,'' {\em Monthly Notices of
  the Royal Astronomical Society: Letters}, vol.~415, pp.~L40--L44, 07 2011.

\bibitem{Kanehisa_2024}
K.~J. Kanehisa, M.~S. Pawlowski, N.~Heesters, and O.~Müller, ``A too-many
  dwarf satellite galaxies problem in the matlas low-to-moderate density
  fields,'' {\em Astronomy amp; Astrophysics}, vol.~686, p.~A280, June 2024.

\bibitem{10.1093/mnras/sts514}
M.~Rocha, A.~H.~G. Peter, J.~S. Bullock, M.~Kaplinghat, S.~Garrison-Kimmel,
  J.~Oñorbe, and L.~A. Moustakas, ``Cosmological simulations with
  self-interacting dark matter – i. constant-density cores and
  substructure,'' {\em Monthly Notices of the Royal Astronomical Society},
  vol.~430, pp.~81--104, 01 2013.

\bibitem{10.1093/mnrasl/sls053}
J.~Zavala, M.~Vogelsberger, and M.~G. Walker, ``Constraining self-interacting
  dark matter with the milky way’s dwarf spheroidals,'' {\em Monthly Notices
  of the Royal Astronomical Society: Letters}, vol.~431, pp.~L20--L24, 02 2013.

\bibitem{10.1093/mnras/sts535}
A.~H.~G. Peter, M.~Rocha, J.~S. Bullock, and M.~Kaplinghat, ``Cosmological
  simulations with self-interacting dark matter – ii. halo shapes versus
  observations,'' {\em Monthly Notices of the Royal Astronomical Society},
  vol.~430, pp.~105--120, 01 2013.

\bibitem{PhysRevLett.84.3760}
D.~N. Spergel and P.~J. Steinhardt, ``Observational evidence for
  self-interacting cold dark matter,'' {\em Phys. Rev. Lett.}, vol.~84,
  pp.~3760--3763, Apr 2000.

\bibitem{ZHANG2023106967}
K.~Zhang, L.-W. Luo, J.-S. Tsao, C.-S. Chen, and F.-L. Lin, ``Dark stars and
  gravitational waves: Topical review,'' {\em Results in Physics}, vol.~53,
  p.~106967, 2023.

\bibitem{Serebrov:2017bzo}
A.~P. Serebrov {\em et~al.}, ``{Neutron lifetime measurements with a large
  gravitational trap for ultracold neutrons},'' {\em Phys. Rev. C}, vol.~97,
  no.~5, p.~055503, 2018.

\bibitem{Pattie:2017vsj}
R.~W. Pattie, Jr. {\em et~al.}, ``{Measurement of the neutron lifetime using a
  magneto-gravitational trap and in situ detection},'' {\em Science}, vol.~360,
  no.~6389, pp.~627--632, 2018.

\bibitem{TAN2019134921}
W.~Tan, ``Neutron oscillations for solving neutron lifetime and dark matter
  puzzles,'' {\em Physics Letters B}, vol.~797, p.~134921, 2019.

\bibitem{PhysRevC.85.065503}
A.~Steyerl, J.~M. Pendlebury, C.~Kaufman, S.~S. Malik, and A.~M. Desai,
  ``Quasielastic scattering in the interaction of ultracold neutrons with a
  liquid wall and application in a reanalysis of the mambo i neutron-lifetime
  experiment,'' {\em Phys. Rev. C}, vol.~85, p.~065503, Jun 2012.

\bibitem{Fornal:2018eol}
B.~Fornal and B.~Grinstein, ``{Dark Matter Interpretation of the Neutron Decay
  Anomaly},'' {\em Phys. Rev. Lett.}, vol.~120, no.~19, p.~191801, 2018.
\newblock [Erratum: Phys.Rev.Lett. 124, 219901 (2020)].

\bibitem{Grinstein:2018ptl}
B.~Grinstein, C.~Kouvaris, and N.~G. Nielsen, ``{Neutron Star Stability in
  Light of the Neutron Decay Anomaly},'' {\em Phys. Rev. Lett.}, vol.~123,
  no.~9, p.~091601, 2019.

\bibitem{Fornal:2020gto}
B.~Fornal and B.~Grinstein, ``{Neutron\textquoteright{}s dark secret},'' {\em
  Mod. Phys. Lett. A}, vol.~35, no.~31, p.~2030019, 2020.

\bibitem{RevModPhys.83.1173}
F.~E. Wietfeldt and G.~L. Greene, ``Colloquium: The neutron lifetime,'' {\em
  Rev. Mod. Phys.}, vol.~83, pp.~1173--1192, Nov 2011.

\bibitem{PhysRevD.65.063006}
L.~Lindblom and B.~J. Owen, ``Effect of hyperon bulk viscosity on neutron-star
  r-modes,'' {\em Phys. Rev. D}, vol.~65, p.~063006, Mar 2002.

\bibitem{2010PhRvC..82b5803J}
T.~{Jha}, H.~{Mishra}, and V.~{Sreekanth}, ``{Bulk viscosity in a hyperonic
  star and r-mode instability},'' {\em prc}, vol.~82, p.~025803, Aug. 2010.

\bibitem{2014CQGra..31o5002V}
C.~{V{\'a}squez Flores} and G.~{Lugones}, ``{Discriminating hadronic and quark
  stars through gravitational waves of fluid pulsation modes},'' {\em Classical
  and Quantum Gravity}, vol.~31, p.~155002, Aug. 2014.

\bibitem{2022PhRvD.106l3002Z}
T.~{Zhao} and J.~M. {Lattimer}, ``{Universal relations for neutron star f -mode
  and g -mode oscillations},'' {\em prd}, vol.~106, p.~123002, Dec. 2022.

\bibitem{Punturo_2010}
M.~Punturo, M.~Abernathy, F.~Acernese, B.~Allen, N.~Andersson, K.~Arun,
  F.~Barone, B.~Barr, M.~Barsuglia, M.~Beker, N.~Beveridge, S.~Birindelli,
  S.~Bose, L.~Bosi, S.~Braccini, C.~Bradaschia, T.~Bulik, E.~Calloni, G.~Cella,
  E.~C. Mottin, S.~Chelkowski, A.~Chincarini, J.~Clark, E.~Coccia, C.~Colacino,
  J.~Colas, A.~Cumming, L.~Cunningham, E.~Cuoco, S.~Danilishin, K.~Danzmann,
  G.~De~Luca, R.~De~Salvo, T.~Dent, R.~De~Rosa, L.~Di~Fiore, A.~Di~Virgilio,
  M.~Doets, V.~Fafone, P.~Falferi, R.~Flaminio, J.~Franc, F.~Frasconi,
  A.~Freise, P.~Fulda, J.~Gair, G.~Gemme, A.~Gennai, A.~Giazotto,
  K.~Glampedakis, M.~Granata, H.~Grote, G.~Guidi, G.~Hammond, M.~Hannam,
  J.~Harms, D.~Heinert, M.~Hendry, I.~Heng, E.~Hennes, S.~Hild, J.~Hough,
  S.~Husa, S.~Huttner, G.~Jones, F.~Khalili, K.~Kokeyama, K.~Kokkotas,
  B.~Krishnan, M.~Lorenzini, H.~Lück, E.~Majorana, I.~Mandel, V.~Mandic,
  I.~Martin, C.~Michel, Y.~Minenkov, N.~Morgado, S.~Mosca, B.~Mours,
  H.~Müller–Ebhardt, P.~Murray, R.~Nawrodt, J.~Nelson, R.~Oshaughnessy,
  C.~D. Ott, C.~Palomba, A.~Paoli, G.~Parguez, A.~Pasqualetti, R.~Passaquieti,
  D.~Passuello, L.~Pinard, R.~Poggiani, P.~Popolizio, M.~Prato, P.~Puppo,
  D.~Rabeling, P.~Rapagnani, J.~Read, T.~Regimbau, H.~Rehbein, S.~Reid,
  L.~Rezzolla, F.~Ricci, F.~Richard, A.~Rocchi, S.~Rowan, A.~Rüdiger,
  B.~Sassolas, B.~Sathyaprakash, R.~Schnabel, C.~Schwarz, P.~Seidel, A.~Sintes,
  K.~Somiya, F.~Speirits, K.~Strain, S.~Strigin, P.~Sutton, S.~Tarabrin,
  A.~Thüring, J.~van~den Brand, C.~van Leewen, M.~van Veggel, C.~van~den
  Broeck, A.~Vecchio, J.~Veitch, F.~Vetrano, A.~Vicere, S.~Vyatchanin,
  B.~Willke, G.~Woan, P.~Wolfango, and K.~Yamamoto, ``The einstein telescope: a
  third-generation gravitational wave observatory,'' {\em Classical and Quantum
  Gravity}, vol.~27, p.~194002, sep 2010.

\bibitem{Abbott_2017}
B.~Abbott, R.~Abbott, T.~Abbott, F.~Acernese, K.~Ackley, C.~Adams, T.~Adams,
  P.~Addesso, R.~Adhikari, V.~Adya, and et~al., ``Gw170817: Observation of
  gravitational waves from a binary neutron star inspiral,'' {\em Physical
  Review Letters}, vol.~119, Oct 2017.

\bibitem{PhysRevD.85.024030}
B.~Zink, P.~D. Lasky, and K.~D. Kokkotas, ``Are gravitational waves from giant
  magnetar flares observable?,'' {\em Phys. Rev. D}, vol.~85, p.~024030, Jan
  2012.

\bibitem{ASCENZI2024102935}
S.~Ascenzi, V.~Graber, and N.~Rea, ``Neutron-star measurements in the
  multi-messenger era,'' {\em Astroparticle Physics}, vol.~158, p.~102935,
  2024.

\bibitem{PAUL2009157}
S.~Paul, ``The puzzle of neutron lifetime,'' {\em Nuclear Instruments and
  Methods in Physics Research Section A: Accelerators, Spectrometers, Detectors
  and Associated Equipment}, vol.~611, no.~2, pp.~157--166, 2009.
\newblock Particle Physics with Slow Neutrons.

\bibitem{PhysRevLett.127.162501}
F.~M. Gonzalez, E.~M. Fries, C.~Cude-Woods, T.~Bailey, M.~Blatnik, L.~J.
  Broussard, N.~B. Callahan, J.~H. Choi, S.~M. Clayton, S.~A. Currie, M.~Dawid,
  E.~B. Dees, B.~W. Filippone, W.~Fox, P.~Geltenbort, E.~George, L.~Hayen,
  K.~P. Hickerson, M.~A. Hoffbauer, K.~Hoffman, A.~T. Holley, T.~M. Ito,
  A.~Komives, C.-Y. Liu, M.~Makela, C.~L. Morris, R.~Musedinovic,
  C.~O'Shaughnessy, R.~W. Pattie, J.~Ramsey, D.~J. Salvat, A.~Saunders, E.~I.
  Sharapov, S.~Slutsky, V.~Su, X.~Sun, C.~Swank, Z.~Tang, W.~Uhrich,
  J.~Vanderwerp, P.~Walstrom, Z.~Wang, W.~Wei, and A.~R. Young, ``Improved
  neutron lifetime measurement with $\mathrm{UCN}\ensuremath{\tau}$,'' {\em
  Phys. Rev. Lett.}, vol.~127, p.~162501, Oct 2021.

\bibitem{10.1093/ptep/ptaa169}
K.~Hirota, G.~Ichikawa, S.~Ieki, T.~Ino, Y.~Iwashita, M.~Kitaguchi,
  R.~Kitahara, J.~Koga, K.~Mishima, T.~Mogi, K.~Morikawa, A.~Morishita,
  N.~Nagakura, H.~Oide, H.~Okabe, H.~Otono, Y.~Seki, D.~Sekiba, T.~Shima, H.~M.
  Shimizu, N.~Sumi, H.~Sumino, T.~Tomita, H.~Uehara, T.~Yamada, S.~Yamashita,
  K.~Yano, M.~Yokohashi, and T.~Yoshioka, ``Neutron lifetime measurement with
  pulsed cold neutrons,'' {\em Progress of Theoretical and Experimental
  Physics}, vol.~2020, p.~123C02, 12 2020.

\bibitem{PhysRevLett.111.222501}
A.~T. Yue, M.~S. Dewey, D.~M. Gilliam, G.~L. Greene, A.~B. Laptev, J.~S. Nico,
  W.~M. Snow, and F.~E. Wietfeldt, ``Improved determination of the neutron
  lifetime,'' {\em Phys. Rev. Lett.}, vol.~111, p.~222501, Nov 2013.

\bibitem{Otono:2016fsv}
H.~Otono, ``{LiNA \textendash{} Lifetime of neutron apparatus with time
  projection chamber and solenoid coil},'' {\em Nucl. Instrum. Meth. A},
  vol.~845, pp.~278--280, 2017.

\bibitem{Olive_2016}
K.~Olive, ``Review of particle physics,'' {\em Chinese Physics C}, vol.~40,
  p.~100001, oct 2016.

\bibitem{UCNt:2021pcg}
F.~M. Gonzalez {\em et~al.}, ``{Improved Neutron Lifetime Measurement with
  UCN\ensuremath{\tau}},'' {\em Phys. Rev. Lett.}, vol.~127, no.~16, p.~162501,
  2021.

\bibitem{doi:10.1126/science.aan8895}
R.~W. Pattie, N.~B. Callahan, C.~Cude-Woods, E.~R. Adamek, L.~J. Broussard,
  S.~M. Clayton, S.~A. Currie, E.~B. Dees, X.~Ding, E.~M. Engel, D.~E. Fellers,
  W.~Fox, P.~Geltenbort, K.~P. Hickerson, M.~A. Hoffbauer, A.~T. Holley,
  A.~Komives, C.-Y. Liu, S.~W.~T. MacDonald, M.~Makela, C.~L. Morris, J.~D.
  Ortiz, J.~Ramsey, D.~J. Salvat, A.~Saunders, S.~J. Seestrom, E.~I. Sharapov,
  S.~K. Sjue, Z.~Tang, J.~Vanderwerp, B.~Vogelaar, P.~L. Walstrom, Z.~Wang,
  W.~Wei, H.~L. Weaver, J.~W. Wexler, T.~L. Womack, A.~R. Young, and B.~A.
  Zeck, ``Measurement of the neutron lifetime using a magneto-gravitational
  trap and in situ detection,'' {\em Science}, vol.~360, no.~6389,
  pp.~627--632, 2018.

\bibitem{PhysRevC.97.055503}
A.~P. Serebrov, E.~A. Kolomensky, A.~K. Fomin, I.~A. Krasnoshchekova, A.~V.
  Vassiljev, D.~M. Prudnikov, I.~V. Shoka, A.~V. Chechkin, M.~E. Chaikovskiy,
  V.~E. Varlamov, S.~N. Ivanov, A.~N. Pirozhkov, P.~Geltenbort, O.~Zimmer,
  T.~Jenke, M.~Van~der Grinten, and M.~Tucker, ``Neutron lifetime measurements
  with a large gravitational trap for ultracold neutrons,'' {\em Phys. Rev. C},
  vol.~97, p.~055503, May 2018.

\bibitem{Tang:2018eln}
Z.~Tang {\em et~al.}, ``{Search for the Neutron Decay n$\rightarrow$ X+$\gamma$
  where X is a dark matter particle},'' {\em Phys. Rev. Lett.}, vol.~121,
  no.~2, p.~022505, 2018.

\bibitem{Serebrov:2007gw}
A.~P. Serebrov {\em et~al.}, ``{Experimental search for neutron: Mirror neutron
  oscillations using storage of ultracold neutrons},'' {\em Phys. Lett. B},
  vol.~663, pp.~181--185, 2008.

\bibitem{Motta:2018rxp}
T.~F. Motta, P.~A.~M. Guichon, and A.~W. Thomas, ``{Implications of Neutron
  Star Properties for the Existence of Light Dark Matter},'' {\em J. Phys. G},
  vol.~45, no.~5, p.~05LT01, 2018.

\bibitem{Motta:2018bil}
T.~F. Motta, P.~A.~M. Guichon, and A.~W. Thomas, ``{Neutron to Dark Matter
  Decay in Neutron Stars},'' {\em Int. J. Mod. Phys. A}, vol.~33, no.~31,
  p.~1844020, 2018.

\bibitem{PhysRevLett.121.061801}
G.~Baym, D.~H. Beck, P.~Geltenbort, and J.~Shelton, ``Testing dark decays of
  baryons in neutron stars,'' {\em Phys. Rev. Lett.}, vol.~121, p.~061801, Aug
  2018.

\bibitem{2018_sa}
D.~McKeen, A.~E. Nelson, S.~Reddy, and D.~Zhou, ``Neutron stars exclude light
  dark baryons,'' {\em Physical Review Letters}, vol.~121, Aug 2018.

\bibitem{Husain:2022bxl}
W.~Husain, T.~F. Motta, and A.~W. Thomas, ``{Consequences of neutron decay
  inside neutron stars},'' {\em JCAP}, vol.~10, p.~028, 2022.

\bibitem{2019ApJ...887L..24M}
M.~{Miller}, F.~{Lamb}, A.~{Dittmann}, S.~{Bogdanov}, Z.~{Arzoumanian},
  K.~{Gendreau}, S.~{Guillot}, A.~{Harding}, W.~{Ho}, J.~{Lattimer},
  R.~{Ludlam}, S.~{Mahmoodifar}, S.~{Morsink}, P.~{Ray}, T.~{Strohmayer},
  K.~{Wood}, T.~{Enoto}, R.~{Foster}, T.~{Okajima}, G.~{Prigozhin}, and
  Y.~{Soong}, ``{PSR J0030+0451 Mass and Radius from NICER Data and
  Implications for the Properties of Neutron Star Matter},'' {\em apjl},
  vol.~887, p.~L24, Dec. 2019.

\bibitem{2019ApJ...887L..21R}
T.~E. {Riley}, A.~{Watts}, S.~{Bogdanov}, P.~{Ray}, R.~{Ludlam}, S.~{Guillot},
  Z.~{Arzoumanian}, C.~{Baker}, A.~{Bilous}, D.~{Chakrabarty}, K.~{Gendreau},
  A.~{Harding}, W.~{Ho}, J.~{Lattimer}, S.~{Morsink}, and T.~{Strohmayer}, ``{A
  NICER View of PSR J0030+0451: Millisecond Pulsar Parameter Estimation},''
  {\em apjl}, vol.~887, p.~L21, Dec. 2019.

\bibitem{2018PhRvL.121p1101A}
B.~P. {Abbott}, R.~{Abbott}, T.~D. {Abbott}, F.~{Acernese}, K.~{Ackley},
  C.~{Adams}, T.~{Adams}, P.~{Addesso}, R.~X. {Adhikari}, V.~B. {Adya},
  C.~{Affeldt}, B.~{Agarwal}, M.~{Agathos}, K.~{Agatsuma}, N.~{Aggarwal}, O.~D.
  {Aguiar}, L.~{Aiello}, A.~{Ain}, P.~{Ajith}, B.~{Allen}, G.~{Allen},
  A.~{Allocca}, M.~A. {Aloy}, P.~A. {Altin}, A.~{Amato}, A.~{Ananyeva}, S.~B.
  {Anderson}, W.~G. {Anderson}, S.~V. {Angelova}, S.~{Antier}, S.~{Appert},
  K.~{Arai}, M.~C. {Araya}, J.~S. {Areeda}, M.~{Ar{\`e}ne}, N.~{Arnaud}, K.~G.
  {Arun}, S.~{Ascenzi}, G.~{Ashton}, M.~{Ast}, S.~M. {Aston}, P.~{Astone},
  D.~V. {Atallah}, F.~{Aubin}, P.~{Aufmuth}, C.~{Aulbert}, K.~{AultONeal},
  C.~{Austin}, A.~{Avila-Alvarez}, S.~{Babak}, P.~{Bacon}, F.~{Badaracco},
  M.~K.~M. {Bader}, S.~{Bae}, P.~T. {Baker}, F.~{Baldaccini}, G.~{Ballardin},
  S.~W. {Ballmer}, S.~{Banagiri}, J.~C. {Barayoga}, S.~E. {Barclay}, B.~C.
  {Barish}, D.~{Barker}, K.~{Barkett}, S.~{Barnum}, F.~{Barone}, B.~{Barr},
  L.~{Barsotti}, M.~{Barsuglia}, D.~{Barta}, J.~{Bartlett}, I.~{Bartos},
  R.~{Bassiri}, A.~{Basti}, J.~C. {Batch}, M.~{Bawaj}, J.~C. {Bayley},
  M.~{Bazzan}, B.~{B{\'e}csy}, C.~{Beer}, M.~{Bejger}, I.~{Belahcene}, A.~S.
  {Bell}, D.~{Beniwal}, M.~{Bensch}, B.~K. {Berger}, G.~{Bergmann},
  S.~{Bernuzzi}, J.~J. {Bero}, C.~P.~L. {Berry}, D.~{Bersanetti},
  A.~{Bertolini}, J.~{Betzwieser}, R.~{Bhandare}, I.~A. {Bilenko}, S.~A.
  {Bilgili}, G.~{Billingsley}, C.~R. {Billman}, J.~{Birch}, R.~{Birney},
  O.~{Birnholtz}, S.~{Biscans}, S.~{Biscoveanu}, A.~{Bisht}, M.~{Bitossi},
  M.~A. {Bizouard}, J.~K. {Blackburn}, J.~{Blackman}, C.~D. {Blair}, D.~G.
  {Blair}, R.~M. {Blair}, S.~{Bloemen}, O.~{Bock}, N.~{Bode}, M.~{Boer},
  Y.~{Boetzel}, G.~{Bogaert}, A.~{Bohe}, F.~{Bondu}, E.~{Bonilla},
  R.~{Bonnand}, P.~{Booker}, B.~A. {Boom}, C.~D. {Booth}, R.~{Bork},
  V.~{Boschi}, S.~{Bose}, K.~{Bossie}, V.~{Bossilkov}, J.~{Bosveld},
  Y.~{Bouffanais}, A.~{Bozzi}, C.~{Bradaschia}, P.~R. {Brady}, A.~{Bramley},
  M.~{Branchesi}, J.~E. {Brau}, T.~{Briant}, F.~{Brighenti}, A.~{Brillet},
  M.~{Brinkmann}, V.~{Brisson}, P.~{Brockill}, A.~F. {Brooks}, D.~D. {Brown},
  S.~{Brunett}, C.~C. {Buchanan}, A.~{Buikema}, T.~{Bulik}, H.~J. {Bulten},
  A.~{Buonanno}, D.~{Buskulic}, C.~{Buy}, R.~L. {Byer}, M.~{Cabero},
  L.~{Cadonati}, G.~{Cagnoli}, C.~{Cahillane}, J.~{Calder{\'o}n Bustillo},
  T.~A. {Callister}, E.~{Calloni}, J.~B. {Camp}, M.~{Canepa}, P.~{Canizares},
  K.~C. {Cannon}, H.~{Cao}, J.~{Cao}, C.~D. {Capano}, E.~{Capocasa},
  F.~{Carbognani}, S.~{Caride}, M.~F. {Carney}, G.~{Carullo}, J.~{Casanueva
  Diaz}, C.~{Casentini}, S.~{Caudill}, M.~{Cavagli{\`a}}, F.~{Cavalier},
  R.~{Cavalieri}, G.~{Cella}, C.~B. {Cepeda}, P.~{Cerd{\'a}-Dur{\'a}n},
  G.~{Cerretani}, E.~{Cesarini}, O.~{Chaibi}, S.~J. {Chamberlin}, M.~{Chan},
  S.~{Chao}, P.~{Charlton}, E.~{Chase}, E.~{Chassande-Mottin}, D.~{Chatterjee},
  K.~{Chatziioannou}, B.~D. {Cheeseboro}, H.~Y. {Chen}, X.~{Chen}, Y.~{Chen},
  H.~P. {Cheng}, H.~Y. {Chia}, and A.~{Chincarini}, ``{GW170817: Measurements
  of Neutron Star Radii and Equation of State},'' {\em prl}, vol.~121,
  p.~161101, Oct. 2018.

\bibitem{2021ApJ...918L..28M}
M.~C. {Miller}, F.~K. {Lamb}, A.~J. {Dittmann}, S.~{Bogdanov},
  Z.~{Arzoumanian}, K.~C. {Gendreau}, S.~{Guillot}, W.~C.~G. {Ho}, J.~M.
  {Lattimer}, M.~{Loewenstein}, S.~M. {Morsink}, P.~S. {Ray}, M.~T. {Wolff},
  C.~L. {Baker}, T.~{Cazeau}, S.~{Manthripragada}, C.~B. {Markwardt},
  T.~{Okajima}, S.~{Pollard}, I.~{Cognard}, H.~T. {Cromartie}, E.~{Fonseca},
  L.~{Guillemot}, M.~{Kerr}, A.~{Parthasarathy}, T.~T. {Pennucci}, S.~{Ransom},
  and I.~{Stairs}, ``{The Radius of PSR J0740+6620 from NICER and XMM-Newton
  Data},'' {\em apjl}, vol.~918, p.~L28, Sept. 2021.

\bibitem{Husain_2022Conseq}
W.~Husain, T.~F. Motta, and A.~W. Thomas, ``Consequences of neutron decay
  inside neutron stars,'' {\em Journal of Cosmology and Astroparticle Physics},
  vol.~2022, p.~028, oct 2022.

\bibitem{BAYM1971225}
G.~Baym, H.~A. Bethe, and C.~J. Pethick, ``Neutron star matter,'' {\em Nuclear
  Physics A}, vol.~175, no.~2, pp.~225 -- 271, 1971.

\bibitem{Lawley:2006ps}
S.~Lawley, W.~Bentz, and A.~W. Thomas, ``{Nucleons, nuclear matter and quark
  matter: A Unified NJL approach},'' {\em J. Phys. G}, vol.~32, pp.~667--680,
  2006.

\bibitem{Whittenbury:2013wma}
D.~L. Whittenbury, J.~D. Carroll, A.~W. Thomas, K.~Tsushima, and J.~R. Stone,
  ``{Quark-Meson Coupling Model, Nuclear Matter Constraints and Neutron Star
  Properties},'' {\em Phys. Rev. C}, vol.~89, p.~065801, 2014.

\bibitem{Whittenbury:2015ziz}
D.~L. Whittenbury, H.~H. Matevosyan, and A.~W. Thomas, ``{Hybrid stars using
  the quark-meson coupling and proper-time Nambu\textendash{}Jona-Lasinio
  models},'' {\em Phys. Rev. C}, vol.~93, no.~3, p.~035807, 2016.

\bibitem{PhysRevD.4.1601}
A.~R. Bodmer, ``Collapsed nuclei,'' {\em Phys. Rev. D}, vol.~4, pp.~1601--1606,
  Sep 1971.

\bibitem{PhysRevD.30.272}
E.~Witten, ``Cosmic separation of phases,'' {\em Phys. Rev. D}, vol.~30,
  pp.~272--285, Jul 1984.

\bibitem{Bombaci_2004}
I.~Bombaci, I.~Parenti, and I.~Vidana, ``Quark deconfinement and implications
  for the radius and the limiting mass of compact stars,'' {\em The
  Astrophysical Journal}, vol.~614, pp.~314--325, oct 2004.

\bibitem{PhysRevD.102.083003}
J.~Ren and C.~Zhang, ``Quantum nucleation of up-down quark matter and
  astrophysical implications,'' {\em Phys. Rev. D}, vol.~102, p.~083003, Oct
  2020.

\bibitem{doi:10.1143/JPSJ.58.3555}
H.~Terazawa, ``Super-hypernuclei in the quark-shell model,'' {\em Journal of
  the Physical Society of Japan}, vol.~58, no.~10, pp.~3555--3563, 1989.

\bibitem{2012_a}
I.~Bednarek, P.~Haensel, J.~L. Zdunik, M.~Bejger, and R.~Mańka, ``Hyperons in
  neutron-star cores and a 2m pulsar,'' {\em Astronomy \& Astrophysics},
  vol.~543, p.~A157, Jul 2012.

\bibitem{doi:10.1063/1.4909561}
I.~Vidaña, ``Hyperons and neutron stars,'' {\em AIP Conference Proceedings},
  vol.~1645, no.~1, pp.~79--85, 2015.

\bibitem{2016_a}
M.~Oertel, F.~Gulminelli, C.~Providência, and A.~R. Raduta, ``Hyperons in
  neutron stars and supernova cores,'' {\em The European Physical Journal A},
  vol.~52, Mar 2016.

\bibitem{PhysRevC.58.1804}
A.~Akmal, V.~R. Pandharipande, and D.~G. Ravenhall, ``Equation of state of
  nucleon matter and neutron star structure,'' {\em Phys. Rev. C}, vol.~58,
  pp.~1804--1828, Sep 1998.

\bibitem{BALBERG1997435}
S.~Balberg and A.~Gal, ``An effective equation of state for dense matter with
  strangeness,'' {\em Nuclear Physics A}, vol.~625, no.~1, pp.~435--472, 1997.

\bibitem{1985ApJ...293..470G}
N.~K. {Glendenning}, ``{Neutron stars are giant hypernuclei ?},'' {\em apj},
  vol.~293, pp.~470--493, June 1985.

\bibitem{KAPLAN198657}
D.~Kaplan and A.~Nelson, ``Strange goings on in dense nucleonic matter,'' {\em
  Physics Letters B}, vol.~175, no.~1, pp.~57--63, 1986.

\bibitem{PhysRevLett.79.1603}
N.~K. Glendenning, S.~Pei, and F.~Weber, ``Signal of quark deconfinement in the
  timing structure of pulsar spin-down,'' {\em Phys. Rev. Lett.}, vol.~79,
  pp.~1603--1606, Sep 1997.

\bibitem{PhysRevLett.67.2414}
N.~K. Glendenning and S.~A. Moszkowski, ``Reconciliation of neutron-star masses
  and binding of the \ensuremath{\Lambda} in hypernuclei,'' {\em Phys. Rev.
  Lett.}, vol.~67, pp.~2414--2417, Oct 1991.

\bibitem{Glendenning1997}
N.~K. Glendenning, {\em Quark Stars}, pp.~289--302.
\newblock New York, NY: Springer US, 1997.

\bibitem{Haensel2017}
P.~Haensel and J.~L. Zdunik, {\em Nuclear Matter in Neutron Stars},
  pp.~1331--1351.
\newblock Cham: Springer International Publishing, 2017.

\bibitem{1980PhR....61...71S}
E.~V. {Shuryak}, ``{Quantum chromodynamics and the theory of superdense
  matter},'' {\em physrep}, vol.~61, pp.~71--158, May 1980.

\bibitem{weber2007neutron}
F.~Weber, R.~Negreiros, and P.~Rosenfield, ``Neutron star interiors and the
  equation of state of superdense matter,'' 2007.

\bibitem{2019_fri}
W.~M. Spinella and F.~Weber, ``Hyperonic neutron star matter in light of
  gw170817,'' {\em Astronomische Nachrichten}, vol.~340, p.~145–150, Jan
  2019.

\bibitem{Weber2016}
F.~Weber, {\em Strange Quark Matter Inside Neutron Stars}, pp.~1--24.
\newblock Cham: Springer International Publishing, 2016.

\bibitem{Terazawa:2001gg}
H.~Terazawa, ``{A new trend in high-energy physics: Current topics in nuclear
  and particle physics},'' in {\em {International Conference on New Trends in
  High-Energy Physics: Experiment, Phenomenology, Theory}}, pp.~246--255, 9
  2001.

\bibitem{Husain_2021}
W.~Husain and A.~W. Thomas, ``Hybrid stars with hyperons and strange quark
  matter,'' {\em PROCEEDINGS OF THE 14TH ASIA-PACIFIC PHYSICS CONFERENCE},
  2021.

\bibitem{2001_lattimer}
J.~M. Lattimer and M.~Prakash, ``Neutron star structure and the equation of
  state,'' {\em The Astrophysical Journal}, vol.~550, p.~426–442, Mar 2001.

\bibitem{2020_latti}
T.~Zhao and J.~M. Lattimer, ``Quarkyonic matter equation of state in
  beta-equilibrium,'' {\em Physical Review D}, vol.~102, Jul 2020.

\bibitem{2021_l}
C.~Drischler, S.~Han, J.~M. Lattimer, M.~Prakash, S.~Reddy, and T.~Zhao,
  ``Limiting masses and radii of neutron stars and their implications,'' {\em
  Physical Review C}, vol.~103, Apr 2021.

\bibitem{Cierniak:2021knt}
M.~Cierniak and D.~Blaschke, ``{Hybrid neutron stars in the mass-radius
  diagram},'' {\em Astron. Nachr.}, vol.~342, no.~5, pp.~819--825, 2021.

\bibitem{Shahrbaf:2022upc}
M.~Shahrbaf, D.~Blaschke, S.~Typel, G.~R. Farrar, and D.~E. Alvarez-Castillo,
  ``{Sexaquark dilemma in neutron stars and its solution by quark
  deconfinement},'' 2 2022.

\bibitem{10.1143/PTP.108.703}
S.~Nishizaki, Y.~Yamamoto, and T.~Takatsuka, ``{Hyperon-Mixed Neutron Star
  Matter and Neutron Stars*)},'' {\em Progress of Theoretical Physics},
  vol.~108, pp.~703--718, 10 2002.

\bibitem{2017_xyz}
Y.~Yamamoto, H.~Togashi, T.~Tamagawa, T.~Furumoto, N.~Yasutake, and T.~A.
  Rijken, ``Neutron-star radii based on realistic nuclear interactions,'' {\em
  Physical Review C}, vol.~96, Dec 2017.

\bibitem{2022_xxyz}
Y.~Yamamoto, N.~Yasutake, and T.~A. Rijken, ``Quark-quark interaction and quark
  matter in neutron stars,'' {\em Physical Review C}, vol.~105, Jan 2022.

\bibitem{Motta:2022nlj}
T.~F. Motta and A.~W. Thomas, ``{The role of baryon structure in neutron
  stars},'' {\em Mod. Phys. Lett. A}, vol.~37, no.~01, p.~2230001, 2022.

\bibitem{Mukhopadhyay_2017}
S.~Mukhopadhyay, D.~Atta, K.~Imam, D.~N. Basu, and C.~Samanta, ``Compact
  bifluid hybrid stars: hadronic matter mixed with self-interacting fermionic
  asymmetric dark matter,'' {\em The European Physical Journal C}, vol.~77, Jul
  2017.

\bibitem{PhysRevD.77.043515}
G.~Bertone and M.~Fairbairn, ``Compact stars as dark matter probes,'' {\em
  Phys. Rev. D}, vol.~77, p.~043515, Feb 2008.

\bibitem{PhysRevD.77.023006}
C.~Kouvaris, ``Wimp annihilation and cooling of neutron stars,'' {\em Phys.
  Rev. D}, vol.~77, p.~023006, Jan 2008.

\bibitem{CIARCELLUTI201119}
P.~Ciarcelluti and F.~Sandin, ``Have neutron stars a dark matter core?,'' {\em
  Physics Letters B}, vol.~695, no.~1, pp.~19--21, 2011.

\bibitem{Sandin_2009}
F.~Sandin and P.~Ciarcelluti, ``Effects of mirror dark matter on neutron
  stars,'' {\em Astroparticle Physics}, vol.~32, p.~278–284, Dec 2009.

\bibitem{Leung:2011zz}
S.~C. Leung, M.~C. Chu, and L.~M. Lin, ``{Dark-matter admixed neutron stars},''
  {\em Phys. Rev. D}, vol.~84, p.~107301, 2011.

\bibitem{Ellis_2018}
J.~Ellis, G.~Hütsi, K.~Kannike, L.~Marzola, M.~Raidal, and V.~Vaskonen, ``Dark
  matter effects on neutron star properties,'' {\em Physical Review D},
  vol.~97, Jun 2018.

\bibitem{bell2020nucleon}
N.~F. Bell, G.~Busoni, T.~F. Motta, S.~Robles, A.~W. Thomas, and M.~Virgato,
  ``Nucleon structure and strong interactions in dark matter capture in neutron
  stars,'' 2020.

\bibitem{2021_w}
W.~Husain and A.~W. Thomas, ``Possible nature of dark matter,'' {\em Journal of
  Cosmology and Astroparticle Physics}, vol.~2021, p.~086, Oct 2021.

\bibitem{2000NuPhB.564..185M}
E.~W. {Mielke} and F.~E. {Schunck}, ``{Boson stars: alternatives to primordial
  black holes?},'' {\em Nuclear Physics B}, vol.~1, pp.~185--203, Jan. 2000.

\bibitem{Blinnikov:1983gh}
S.~I. Blinnikov and M.~Khlopov, ``{Possible astronomical effects of mirror
  particles},'' {\em Sov. Astron.}, vol.~27, pp.~371--375, 1983.

\bibitem{2019_reddy}
C.~Horowitz and S.~Reddy, ``Gravitational waves from compact dark objects in
  neutron stars,'' {\em Physical Review Letters}, vol.~122, Feb 2019.

\bibitem{2013_red}
B.~Bertoni, A.~E. Nelson, and S.~Reddy, ``Dark matter thermalization in neutron
  stars,'' {\em Physical Review D}, vol.~88, Dec 2013.

\bibitem{berryman2022neutron}
J.~M. Berryman, S.~Gardner, and M.~Zakeri, ``Neutron stars with baryon number
  violation, probing dark sectors,'' 2022.

\bibitem{McKeen:2021jbh}
D.~McKeen, M.~Pospelov, and N.~Raj, ``{Neutron Star Internal Heating
  Constraints on Mirror Matter},'' {\em Phys. Rev. Lett.}, vol.~127, no.~6,
  p.~061805, 2021.

\bibitem{deLavallaz:2010wp}
A.~de~Lavallaz and M.~Fairbairn, ``{Neutron Stars as Dark Matter Probes},''
  {\em Phys. Rev. D}, vol.~81, p.~123521, 2010.

\bibitem{Busoni:2021zoe}
G.~Busoni, ``{Capture of Dark Matter in Neutron Stars},'' 12 2021.

\bibitem{Sen:2021wev}
D.~Sen and A.~Guha, ``{Implications of feebly interacting dark sector on
  neutron star properties and constraints from GW170817},'' {\em Mon. Not. Roy.
  Astron. Soc.}, vol.~504, no.~3, p.~3, 2021.

\bibitem{Guha:2021njn}
A.~Guha and D.~Sen, ``{Feeble DM-SM interaction via new scalar and vector
  mediators in rotating neutron stars},'' {\em JCAP}, vol.~09, p.~027, 2021.

\bibitem{Guichon:1987jp}
P.~A.~M. Guichon, ``{A Possible Quark Mechanism for the Saturation of Nuclear
  Matter},'' {\em Phys. Lett. B}, vol.~200, pp.~235--240, 1988.

\bibitem{Guichon:1995ue}
P.~A.~M. Guichon, K.~Saito, E.~N. Rodionov, and A.~W. Thomas, ``{The Role of
  nucleon structure in finite nuclei},'' {\em Nucl. Phys. A}, vol.~601,
  pp.~349--379, 1996.

\bibitem{Stone:2016qmi}
J.~R. Stone, P.~A.~M. Guichon, P.~G. Reinhard, and A.~W. Thomas, ``{Finite
  Nuclei in the Quark-Meson Coupling Model},'' {\em Phys. Rev. Lett.},
  vol.~116, no.~9, p.~092501, 2016.

\bibitem{RIKOVSKASTONE2007341}
J.~{Rikovska Stone}, P.~Guichon, H.~Matevosyan, and A.~Thomas, ``Cold uniform
  matter and neutron stars in the quark–meson-coupling model,'' {\em Nuclear
  Physics A}, vol.~792, no.~3, pp.~341--369, 2007.

\bibitem{Husain_2020}
W.~Husain and A.~W. Thomas, ``Significance of lower energy density region of
  neutron star and universalities among neutron star properties,'' {\em Journal
  of Physics: Conference Series}, vol.~1643, p.~012066, dec 2020.

\bibitem{Guichon:2018uew}
P.~A.~M. Guichon, J.~R. Stone, and A.~W. Thomas,
  ``{Quark\textendash{}Meson-Coupling (QMC) model for finite nuclei, nuclear
  matter and beyond},'' {\em Prog. Part. Nucl. Phys.}, vol.~100, pp.~262--297,
  2018.

\bibitem{DeGrand:1975cf}
T.~A. DeGrand, R.~L. Jaffe, K.~Johnson, and J.~E. Kiskis, ``{Masses and Other
  Parameters of the Light Hadrons},'' {\em Phys. Rev. D}, vol.~12, p.~2060,
  1975.

\bibitem{Motta:2019tjc}
T.~F. Motta, A.~M. Kalaitzis, S.~Anti\'c, P.~A.~M. Guichon, J.~R. Stone, and
  A.~W. Thomas, ``{Isovector Effects in Neutron Stars, Radii and the GW170817
  Constraint},'' {\em Astrophys. J.}, vol.~878, no.~2, p.~159, 2019.

\bibitem{Urbanec_2013}
M.~Urbanec, J.~C. Miller, and Z.~Stuchlík, ``Quadrupole moments of rotating
  neutron stars and strange stars,'' {\em Monthly Notices of the Royal
  Astronomical Society}, vol.~433, p.~1903–1909, June 2013.

\bibitem{Tolman169}
R.~C. Tolman, ``Effect of inhomogeneity on cosmological models,'' {\em
  Proceedings of the National Academy of Sciences}, vol.~20, no.~3,
  pp.~169--176, 1934.

\bibitem{PhysRev.55.374}
J.~R. Oppenheimer and G.~M. Volkoff, ``On massive neutron cores,'' {\em Phys.
  Rev.}, vol.~55, pp.~374--381, Feb 1939.

\bibitem{1983ApJS...53...73L}
L.~{Lindblom} and S.~L. {Detweiler}, ``{The quadrupole oscillations of neutron
  stars.},'' {\em apjs}, vol.~53, pp.~73--92, Sept. 1983.

\bibitem{1985ApJ...292...12D}
S.~{Detweiler} and L.~{Lindblom}, ``{On the nonradial pulsations of general
  relativistic stellar models},'' {\em apj}, vol.~292, pp.~12--15, May 1985.

\bibitem{Jaiswal_2021}
S.~Jaiswal and D.~Chatterjee, ``Constraining dense matter physics using f-mode
  oscillations in neutron stars,'' {\em Physics}, vol.~3, p.~302–319, May
  2021.

\bibitem{Celato_2025}
M.~Celato, C.~J. Krüger, and K.~D. Kokkotas, ``Probing dark star parameters
  through <mml:math xmlns:mml="http://www.w3.org/1998/math/mathml"
  display="inline"><mml:mi>f</mml:mi></mml:math> -mode gravitational wave
  signals,'' {\em Physical Review D}, vol.~111, Jan. 2025.

\bibitem{Guha_Roy_2024}
D.~Guha~Roy, T.~Malik, S.~Bhattacharya, and S.~Banik, ``Analysis of neutron
  star f-mode oscillations in general relativity with spectral representation
  of nuclear equations of state,'' {\em The Astrophysical Journal}, vol.~968,
  p.~124, jun 2024.

\bibitem{Ozel:2016oaf}
F.~\"Ozel and P.~Freire, ``{Masses, Radii, and the Equation of State of Neutron
  Stars},'' {\em Ann. Rev. Astron. Astrophys.}, vol.~54, pp.~401--440, 2016.

\bibitem{2010Natur.467.1081D}
P.~B. {Demorest}, T.~{Pennucci}, S.~M. {Ransom}, M.~S.~E. {Roberts}, and
  J.~W.~T. {Hessels}, ``{A two-solar-mass neutron star measured using Shapiro
  delay},'' {\em nat}, vol.~467, pp.~1081--1083, Oct. 2010.

\bibitem{2013Sci...340..448A}
J.~{Antoniadis}, P.~C.~C. {Freire}, N.~{Wex}, T.~M. {Tauris}, R.~S. {Lynch},
  M.~H. {van Kerkwijk}, M.~{Kramer}, C.~{Bassa}, V.~S. {Dhillon}, T.~{Driebe},
  J.~W.~T. {Hessels}, V.~M. {Kaspi}, V.~I. {Kondratiev}, N.~{Langer}, T.~R.
  {Marsh}, M.~A. {McLaughlin}, T.~T. {Pennucci}, S.~M. {Ransom}, I.~H.
  {Stairs}, J.~{van Leeuwen}, J.~P.~W. {Verbiest}, and D.~G. {Whelan}, ``{A
  Massive Pulsar in a Compact Relativistic Binary},'' {\em Science}, vol.~340,
  p.~448, Apr. 2013.

\bibitem{dey2024fmodeoscillationsdarkmatter}
D.~Dey, J.~A. Pattnaik, R.~N. Panda, M.~Bhuyan, and S.~K. Patra, ``f-mode
  oscillations of dark matter admixed quarkyonic neutron star,'' 2024.

\bibitem{PhysRevD.110.063025}
S.~Shirke, B.~K. Pradhan, D.~Chatterjee, L.~Sagunski, and J.~Schaffner-Bielich,
  ``Effects of dark matter on $f$-mode oscillations of neutron stars,'' {\em
  Phys. Rev. D}, vol.~110, p.~063025, Sep 2024.

\bibitem{Jyothilakshmi2025}
O.~P. Jyothilakshmi, L.~J. Naik, D.~Sen, A.~Guha, and V.~Sreekanth, ``Effects
  of dark boson mediated feeble interaction between dark matter (dm) and quark
  matter on f-mode oscillation of dm admixed quark stars,'' {\em The European
  Physical Journal C}, vol.~85, no.~461, 2025.

\bibitem{Andersson1998}
N.~Andersson and K.~D. Kokkotas, ``Towards gravitational wave
  asteroseismology,'' {\em Mon. Not. R. Astron. Soc.}, vol.~299,
  pp.~1059--1068, 1998.

\bibitem{10.1046/j.1365-8711.1998.01840.x}
N.~Andersson and K.~D. Kokkotas, ``Towards gravitational wave
  asteroseismology,'' {\em Monthly Notices of the Royal Astronomical Society},
  vol.~299, pp.~1059--1068, 10 1998.

\bibitem{Lioutas2018}
G.~Lioutas and N.~Stergioulas, ``Universal and approximate relations for the
  gravitational-wave damping timescale of f-modes in neutron stars,'' {\em
  General Relativity and Gravitation}, vol.~50, no.~12, 2018.

\bibitem{Lau2010}
H.~K. Lau, P.~T. Leung, and L.~M. Lin, ``Inferring physical parameters of
  compact stars from their f-mode gravitational wave signals,'' {\em The
  Astrophysical Journal}, vol.~714, no.~2, p.~1234, 2010.

\bibitem{Hild_2011}
S.~Hild, M.~Abernathy, F.~Acernese, P.~Amaro-Seoane, N.~Andersson, K.~Arun,
  F.~Barone, B.~Barr, M.~Barsuglia, M.~Beker, N.~Beveridge, S.~Birindelli,
  S.~Bose, L.~Bosi, S.~Braccini, C.~Bradaschia, T.~Bulik, E.~Calloni, G.~Cella,
  E.~C. Mottin, S.~Chelkowski, A.~Chincarini, J.~Clark, E.~Coccia, C.~Colacino,
  J.~Colas, A.~Cumming, L.~Cunningham, E.~Cuoco, S.~Danilishin, K.~Danzmann,
  R.~De~Salvo, T.~Dent, R.~De~Rosa, L.~Di~Fiore, A.~Di~Virgilio, M.~Doets,
  V.~Fafone, P.~Falferi, R.~Flaminio, J.~Franc, F.~Frasconi, A.~Freise,
  D.~Friedrich, P.~Fulda, J.~Gair, G.~Gemme, E.~Genin, A.~Gennai, A.~Giazotto,
  K.~Glampedakis, C.~Gräf, M.~Granata, H.~Grote, G.~Guidi, A.~Gurkovsky,
  G.~Hammond, M.~Hannam, J.~Harms, D.~Heinert, M.~Hendry, I.~Heng, E.~Hennes,
  J.~Hough, S.~Husa, S.~Huttner, G.~Jones, F.~Khalili, K.~Kokeyama,
  K.~Kokkotas, B.~Krishnan, T.~G.~F. Li, M.~Lorenzini, H.~Lück, E.~Majorana,
  I.~Mandel, V.~Mandic, M.~Mantovani, I.~Martin, C.~Michel, Y.~Minenkov,
  N.~Morgado, S.~Mosca, B.~Mours, H.~Müller–Ebhardt, P.~Murray, R.~Nawrodt,
  J.~Nelson, R.~Oshaughnessy, C.~D. Ott, C.~Palomba, A.~Paoli, G.~Parguez,
  A.~Pasqualetti, R.~Passaquieti, D.~Passuello, L.~Pinard, W.~Plastino,
  R.~Poggiani, P.~Popolizio, M.~Prato, M.~Punturo, P.~Puppo, D.~Rabeling,
  P.~Rapagnani, J.~Read, T.~Regimbau, H.~Rehbein, S.~Reid, F.~Ricci,
  F.~Richard, A.~Rocchi, S.~Rowan, A.~Rüdiger, L.~Santamaría, B.~Sassolas,
  B.~Sathyaprakash, R.~Schnabel, C.~Schwarz, P.~Seidel, A.~Sintes, K.~Somiya,
  F.~Speirits, K.~Strain, S.~Strigin, P.~Sutton, S.~Tarabrin, A.~Thüring,
  J.~van~den Brand, M.~van Veggel, C.~van~den Broeck, A.~Vecchio, J.~Veitch,
  F.~Vetrano, A.~Vicere, S.~Vyatchanin, B.~Willke, G.~Woan, and K.~Yamamoto,
  ``Sensitivity studies for third-generation gravitational wave
  observatories,'' {\em Classical and Quantum Gravity}, vol.~28, p.~094013,
  Apr. 2011.

\bibitem{ETsensitivities}
M.~Punturo, ``Et sensitivities page.''
  \url{https://www.et-gw.eu/index.php/etsensitivities}, 2009.
\newblock Last updated: 14 August 2021. Accessed: 26 September 2025.

\bibitem{zheng2025fmodeoscillationsprotoneutronstars}
Z.-Y. Zheng, T.~ting Sun, H.~Chen, J.-B. Wei, X.-P. Zheng, G.~F. Burgio, and
  H.~J. Schulze, ``$f$-mode oscillations of protoneutron stars,'' 2025.

\bibitem{Li_2025}
K.~J. Li, J.~S. Long, K.~Wu, and A.~K.~H. Kong, ``A multimessenger mass
  determination method for lisa neutron star–white dwarf binaries,'' {\em The
  Astrophysical Journal}, vol.~984, p.~41, apr 2025.

\end{thebibliography}
\bibliographystyle{ieeetr}

%\appendix
\end{document}